\documentclass[preprint,10pt,authoryear, 3p]{article}

\usepackage{arxiv}
\usepackage[english]{babel}
\usepackage[utf8x]{inputenc}
\usepackage{amsmath}
\usepackage{graphicx}
\usepackage[colorinlistoftodos]{todonotes}
\usepackage{comment}
\usepackage{color,soul} 
\usepackage{natbib}
\usepackage{makecell}

\title{\textbf{A Computational Model of Representation Learning in the Brain Cortex, Integrating Unsupervised and Reinforcement Learning}}

\author{
    Giovanni Granato*\\
    Laboratory of Computational Embodied Neuroscience\\
    Institute of Cognitive Sciences and Technologies\\
    National Research Council of Italy,\\
    Rome, Italy\\
    School of Computing, Electronics and Mathematics\\ 
    University of Plymouth\\ 
    Plymouth, U.K.\\
    \texttt{giovanni.granato@istc.cnr.it} \\
    \and
    Emilio Cartoni\\
    Laboratory of Computational Embodied Neuroscience\\
    Institute of Cognitive Sciences and Technologies\\
    National Research Council of Italy\\
    Rome, Italy\\
    \texttt{emilio.cartoni@istc.cnr.it} \\
    \and
    Federico Da Rold\\
    Body Action Language Lab\\
    Institute of Cognitive Sciences and Technologies\\
    National Research Council of Italy\\
    Rome, Italy\\
    \texttt{federico.darold@istc.cnr.it} \\
    \and
    Andrea Mattera \\
    Laboratory of Computational Embodied Neuroscience\\
    Institute of Cognitive Sciences and Technologies\\
    National Research Council of Italy\\
    Rome, Italy\\
    \texttt{andrea.mattera@istc.cnr.it} \\
    \and
    Gianluca Baldassarre \\
    Laboratory of Computational Embodied Neuroscience\\
    Institute of Cognitive Sciences and Technologies\\
    National Research Council of Italy\\
    Rome, Italy\\
    \texttt{gianluca.baldassarre@istc.cnr.it} \\
}

\begin{document}

\maketitle

\begin{abstract}
A common view on the brain learning processes proposes that the three classic learning paradigms---unsupervised, reinforcement, and supervised---take place in respectively the cortex, the basal-ganglia, and the cerebellum.
However, dopamine outbursts, usually assumed to encode reward, are not limited to the basal ganglia but also reach prefrontal, motor, and higher sensory cortices. 
We propose that in the cortex the same reward-based trial-and-error processes might support not only the acquisition of motor representations but also of sensory representations.
In particular, reward signals might guide trial-and-error processes that mix with associative learning processes to support the acquisition of representations better serving downstream action selection.
We tested the soundness of this hypothesis with a computational model that integrates unsupervised learning (Contrastive Divergence) and reinforcement learning (REINFORCE).
The model was tested with a task requiring different responses to different visual images grouped in categories involving either colour, shape, or size.
Results show that a balanced mix of unsupervised and reinforcement learning processes leads to the best performance.
%Moreover, the learned category representations tend to both reflect the general statistical properties of the images and to be more disentangled in favour of downstream action selection (action-oriented disentanglement).
Indeed, excessive unsupervised learning tends to under-represent task-relevant features while excessive reinforcement learning tends to initially learn slowly and then to incur in local minima.
These results stimulate future empirical studies on category learning directed to investigate similar effects in the extrastriate visual cortices.
Moreover, they prompt further computational investigations directed to study the possible advantages of integrating unsupervised and reinforcement learning processes.
\end{abstract}

% \keywords{Category learning \and Unsupervised learning \and Reinforcement learning \and Generative models \and Contrastive Divergence \and REINFORCE}

\twocolumn
\section{Introduction}

A classic computational view differentiates the brain learning processes between unsupervised learning (UL) taking place in the cortex, reinforcement learning (RL) based on dopamine dynamics within the basal ganglia, and supervised learning in the cerebellum \citep{Doya1999WhatAretheComputationsoftheCerebellumtheBasalGangliaandtheCerebralCortex,Doya2000ComplementaryRolesofBasalGangliaandCerebelluminLearningandMotorControl}. 
%Fede: sopra ho tolto delle virgole e l'acronimo SL perche' poi non lo usiamo mai nel testo
The biological literature \citep{HoukDavidsBeiser1995ModelsofInformationProcessingintheBasalGanglia,RedgraveGurney2006TheShortLatencyDopamineSignalaRoleinDiscoveringNovelActions}, supported by reinforcement-learning computational models \citep{SuttonBarto2018ReinforcementLearningAnIntroduction,FioreSperatiMannellaMirolliGurneyFirstonDolanBaldassarre2014Keepfocussingstriataldopaminemultiplefunctionsresolvedinasinglemechanismtestedinasimulatedhumanoidrobot}, has shown how
%fede: ho tolto indeed perche' lo usiamo molto spesso. Meglio snellire qui non era proprio necessario secondo me
basal-ganglia learning is strongly driven by reward signals encoded by the dopamine released by the neuromodulator mesolimbic system.
%Focusing on the first two, the biological literature has extensively shown how indeed basal-ganglia learning is strongly driven by reward signals encoded by the dopamine released by the dopamine mesolimbic system \citep{HoukDavidsBeiser1995ModelsofInformationProcessingintheBasalGanglia,RedgraveGurney2006TheShortLatencyDopamineSignalaRoleinDiscoveringNovelActions} and is also supported by reinforcement learning computational models in one of the most successful synergies between neuroscience and modelling \citep{SuttonBarto2018ReinforcementLearningAnIntroduction,FioreSperatiMannellaMirolliGurneyFirstonDolanBaldassarre2014Keepfocussingstriataldopaminemultiplefunctionsresolvedinasinglemechanismtestedinasimulatedhumanoidrobot}.
A similarly extensive literature shows how cortical learning is largely based on associative learning
%processes, in particular taking the form of spike-timing dependent plasticity sensitive to the timing of the spike events to associate 
\citep{MarkramGerstnerSjostrom2011AHistoryofSpikeTimingDependentPlasticity,CaporaleDan2008SpikeTimingDependentPlasticityaHebbianLearningRule}, as also operationalised with computational models \citep{Hopfield1982NeuralNetworksandPhysicalSystemswithEmergentCollectiveComputationalAbilities,GerstnerKistler2002SpikingNeuronModelsSingleNeuronsPopulationsPlasticity,ZappacostaMannellaMirolliBaldassarre2018GeneralDifferentialHebbianLearningCapturingTemporalRelationsbetweenEventsinNeuralNetworksandtheBrain}. 

%But: domapine reaches virtually the whole cortex
However, empirical evidence shows that dopamine also \textit{directly} innervates
the cortex through the neuromodulator mesocortical system having decreasing mediodorsal and anterior-posterior projection gradients spanning far beyond motor cortices \citep{ WilliamsGoldmanRakic1993CharacterizationoftheDopaminergicInnervationofthePrimateFrontalCortexUsingaDopamineSpecificAntibody,jacob2018monoaminergic, niu2020receptor,froudist2020dopamine}.
%impieri2019receptor
Dopamine-based reward signals thus play an important role not only for basal-ganglia but also for the cortex \citep{Wise2004DopamineLearningandMotivation}.
For example, different prefrontal and motor cortices encode different aspects related to reward, such as rewarded outcomes, stimuli associated with these outcomes, actions leading to them, and working memory upload-download
\citep{OReilly2006Biologicallybasedcomputationalmodelsofhighlevelcognition,RushworthNoonanBoormanWaltonBehrens2011FrontalCortexandRewardGuidedLearningandDecisionMaking,MannellaBaldassarre2015Selectionofcorticaldynamicsformotorbehaviourbythebasalganglia,ZeithamovaMackBraunlichDavisSegerKesterenWutz2019BrainMechanismsofConceptLearning}.

Recently we have extended \citep{caligiore2019super} the previous view on the brain learning processes \citep{Doya1999WhatAretheComputationsoftheCerebellumtheBasalGangliaandtheCerebralCortex,Doya2000ComplementaryRolesofBasalGangliaandCerebelluminLearningandMotorControl}
by 
%theoretically FEDE: io lo toglierei per la scorrevolezza, a meno che non si voglia proprio sottolineare che e' un claim teorico. 
proposing that each of the different macro-systems of the brain---the basal ganglia, the cortex, and the cerebellum---involve multiple learning mechanisms.
In this view, cortical plasticity involves both associative (unsupervised) learning processes and trial-and-error (reinforcement) learning processes based on dopamine.
However, to our knowledge how these two learning processes might integrate and affect the learned internal representations of stimuli has not been investigated with computational models.

%Objective of this work
The goal of this work is to propose a hypothesis stating that reward-based trial-and-error processes might lead not only to the acquisition of motor representations, as it is usually assumed, but also of sensory representations.
%FEDE: l'ho tagliata. Mi piace l'incisiva che dice come di assunto di solito perche' ferma il lettore. Il fatto che la operazionalizzamo lo diciamo poche righe sotto. Secondo me qui stiamo con l'ipotesi
In particular, the hypothesis we propose is that:
(a) within cortices, reward-based trial-and-error learning processes and associative learning processes are mixed;
(b)
%1
%non-motor representations are acquired through exploratory noise and the reward-based fixation of the effective solutions found, as those that support the acquisition of actions;
%2
%the same mechanisms support the acquisition of non-motor and motor representations, namely, exploratory noise and the reward-based fixation of the effective solutions found
%3
the trial-and-error mechanisms learning non-motor representations are analogous to those learning motor representations, consisting of exploratory noise and the reward-based fixation of the found effective solutions; 
%4
%these trial-and-error processes are based on mechanisms analogous to those leading to the acquisition of action representations, namely, exploratory noise and the reward-based fixation of the effective solutions found;
%5
%these trial-and-error processes are based on mechanisms analogous to those leading to the acquisition of action representations, namely, exploratory noise and the reward-based fixation of the effective solutions found;
(c) the joint effect of these associative and trial-and-error learning processes leads to the acquisition of action-oriented representations better serving downstream action selection processes.

%Model key elements
To operationalise the hypothesis, we propose a computational model having an overall actor-critic architecture \citep{SuttonBarto2018ReinforcementLearningAnIntroduction} and based on a generative model \citep{BengioGoodfellowCourville2017DeepLearning}.
The generative model is a Deep Belief Network based on two stacked Restricted Boltzmann Machines, \citep{Hinton2002Trainingproductsofexpertsbyminimizingcontrastivedivergence,hinton2006fast} integrating the UL \textit{Contrastive Divergence} \citep{Hinton2002Trainingproductsofexpertsbyminimizingcontrastivedivergence} and RL, using the \textit{REINFORCE} algorithm \citep{williams1992simple} .

%OLD DESCRIPTION OF THE MODEL
%Model architecture
%The model is formed by three key components:
%(a) a \textit{perceptual component} based on a  deep generative neural network (a Deep Belief Network, based on two stacked Restricted Boltzmann Machines, \citealp{Hinton2002Trainingproductsofexpertsbyminimizingcontrastivedivergence,hinton2006fast}) trained with a novel algorithm that balances unsupervised learning (Contrastive Divergence) and reinforcement learning (REINFORCE ) to acquire the representations of visual inputs;
%(b) a \textit{motor component}, based on a neural network that learns through reinforcement learning (Williams reinforcement learning algorithm), that produces an output on the basis of the acquired representations of the world; 
%(c) a \textit{motivational component} that monitors the model performance and produces a reward signal that influences the learning of the two first components.
%
%The last two components form an actor-critic reinforcement learning architecture \citep{sutton1998reinforcement} where the \textit{actor} corresponds to the motor component and the \textit{critic} to the motivational component.

%Mechanisms
The model has an architecture that is different from those commonly used in reinforcement-learning neural network models \citep{MnihKavukcuogluSilverRusuVenessBellemareGravesRiedmillerFidjelandOstrovskiPetersenBeattieSadikAntonoglouKingKumaranWierstraLeggHassabis2015Humanlevelcontrolthroughdeepreinforcementlearning,ArulkumaranDeisenrothBrundageBharath2017DeepReinforcementLearningaBriefSurvey,NguyenNguyenNahavandi2020DeepReinforcementLearningforMultiagentSystemsaReviewofChallengesSolutionsandApplications,ShaoTangZhuLiZhao2019ASurveyofDeepReinforcementLearninginVideoGames}.
% and in models that mix supervised and reinforcement learning \cite{to find}.
In these systems, 
(a) the RL algorithms and reward signals are used to train only the output `motor' layers and 
(b) the effects of reward on inner layers are caused by error gradients that back-propagate from output layers.
On the contrary, we propose here that in the cortex the reward signals \textit{directly} affect the representations formed within the inner layers.

%Model tests
%fede: la riformulerei
% The model is tested with a sorting task that requires the performance of different actions in response to different attributes (e.g., green/blue/red/yellow colours) and categories (colour, shape, or size) of simple geometric images.
%secondo me sta cosa del categories and attributes confonde parecchio. Metterei un generico property oppure category. Ma secondo me gli oggetti hanno proprieta', le categorie sono in testa. Oppure nella relazione agente-oggetto.
The model is tested with sorting tasks requiring the execution of consistent actions in response to different properties (colour, shape, or size) of simple geometric images.
%
% Task we face
This task is inspired to sorting tasks used in the research on category learning \citep{HananiaSmith2010SelectiveAttentionandAttentionSwitchingTowardsaUnifiedDevelopmentalApproach}
where the participants have to sort cards with simple shapes by putting them onto other cards showing similar attributes (e.g., red) according to a certain category (e.g., colour).
%
%babies \citep{taffoni2019motor} requiring the execution of different actions linked to one visual property of objects.
%This task probes both the capacity of learning the production of a task-oriented external output (an abstract version of motor skills) and the acquisition of sensory representations (an abstract version of perceptual skills).
%
%ANTICIPATION OF RESULTS, AND THEIR MEANING.

% Results
The results show that a balanced mix of UL and RL processes leads to higher performance.
Moreover, the learned representations exhibit a category-based \textit{action-oriented disentanglement} effect for which the encoding encompasses both intrinsic statistical regularities and action-relevant visual features of images.
%
%Closure
These results corroborate the hypothesis for which reward-based trial-and-error processes can directly affect sensory representations in the cortex thus tuning them towards action. 

\section{Methods}

%una cosa sulle figure, ricordiamocelo: sono fatte bene ma la figura 2 stampata bianco e nero rende idistinguibili colori delle components. Secondo me possiamo usare linee diverse nelle shapes (rettangoli e trapezi), e.g. motivational solid, motor dashed etc. Cosi' anche in b&n si distinguono 

\subsection{Task and experimental conditions}

The task we used to test the model is based on category learning tasks requiring the production of a response on the basis of specific visual features of stimuli such as colour, shape, and size (see \citealp{ashby2005human, ashby2011human} for an extended analysis of these tasks). 

\begin{figure*}[htb!]
    \centering
	\includegraphics[width=\textwidth]{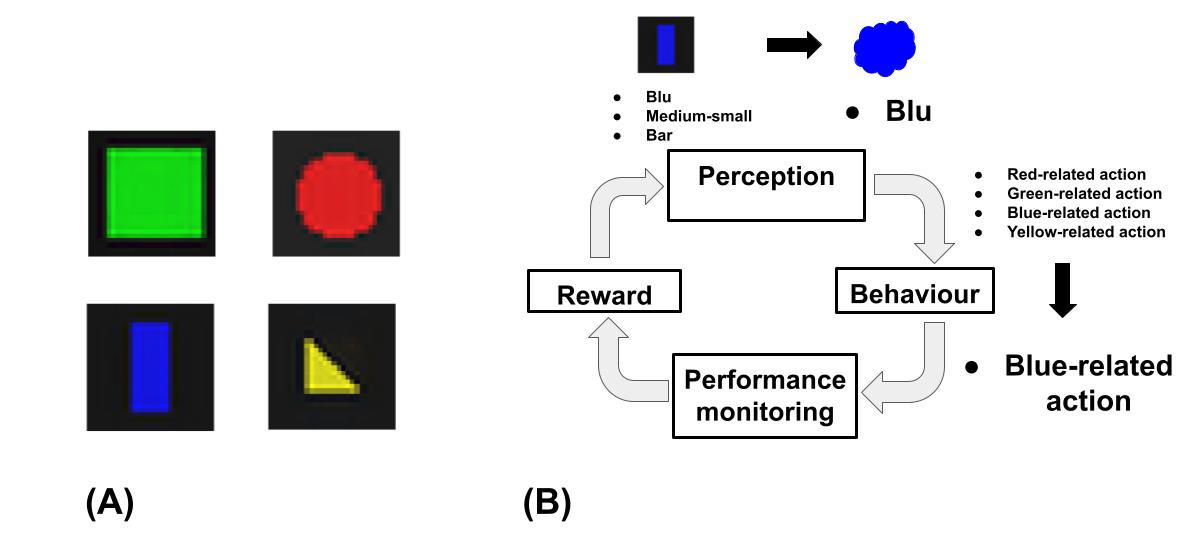}
	 \caption{
	 (A) Examples of the 64 geometrical shapes (circles, squares, parallelepipeds, triangles) used to produce the images. Each image encompasses a different attribute out of the four attributes of each of the three categories colour, shape, and size.
	 (B) A schema of the main model processes involved in its interaction with the environment.
	 }
	 \label{Figure:Task}
\end{figure*}

In particular, we focused on a sub-class of these tasks in which a classification rule is fixed and the participant has to execute a motor action on the basis of the features of a card \citep{HananiaSmith2010SelectiveAttentionandAttentionSwitchingTowardsaUnifiedDevelopmentalApproach}.
%objects features (see \citealp{taffoni2019motor} for an example).
%
The task uses a series of 2D input images of geometrical shapes varying in colour, shape, and size, for example as those shown in Figure~\ref{Figure:Task}A. 
For example in the case of a colour classification rule, the agent should learn to respond with a different output to the different colours (red, green, blue, yellow), hence ignoring the shape and the size.
Figure~\ref{Figure:Task}B summarises the main processes performed by the system during the task performance: perception of the input, behavioural response, performance monitoring, and processing of the reward.
The task was repeated for all the three classification rules involving colour, shape, and size.

\subsection{The architecture of the model and its biological underpinning}

\begin{figure}[htb!]
    \centering
	\includegraphics[width=0.5\textwidth]{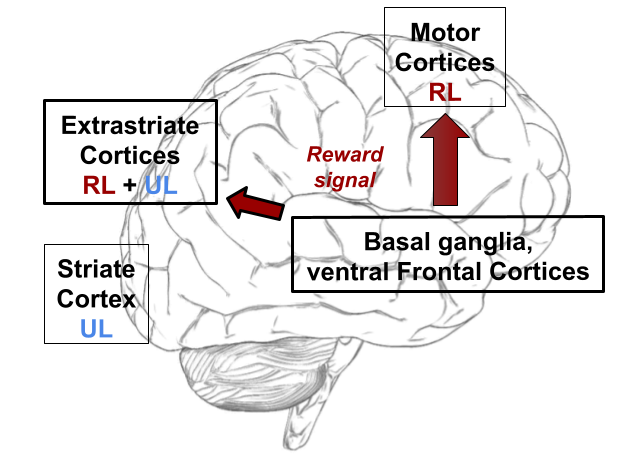}
	 \caption{Scheme of learning processes and targeted brain areas that are addressed by the hypothesis and computational model presented here.
	 The non-motor cortex undergoes both associative learning (UL) and trial-and-error learning (RL). The latter presents a gradient having a decreasing strength moving from the motor cortex towards the striate cortex.
	 The model studies the effects that the mix of unsupervised and reinforcement-learning processes have in extrastriate cortices.}
	 \label{Figure:Brain_learning}
\end{figure}

\begin{figure*}[htb!]
    \centering
    \centering
	\includegraphics[width=\textwidth]{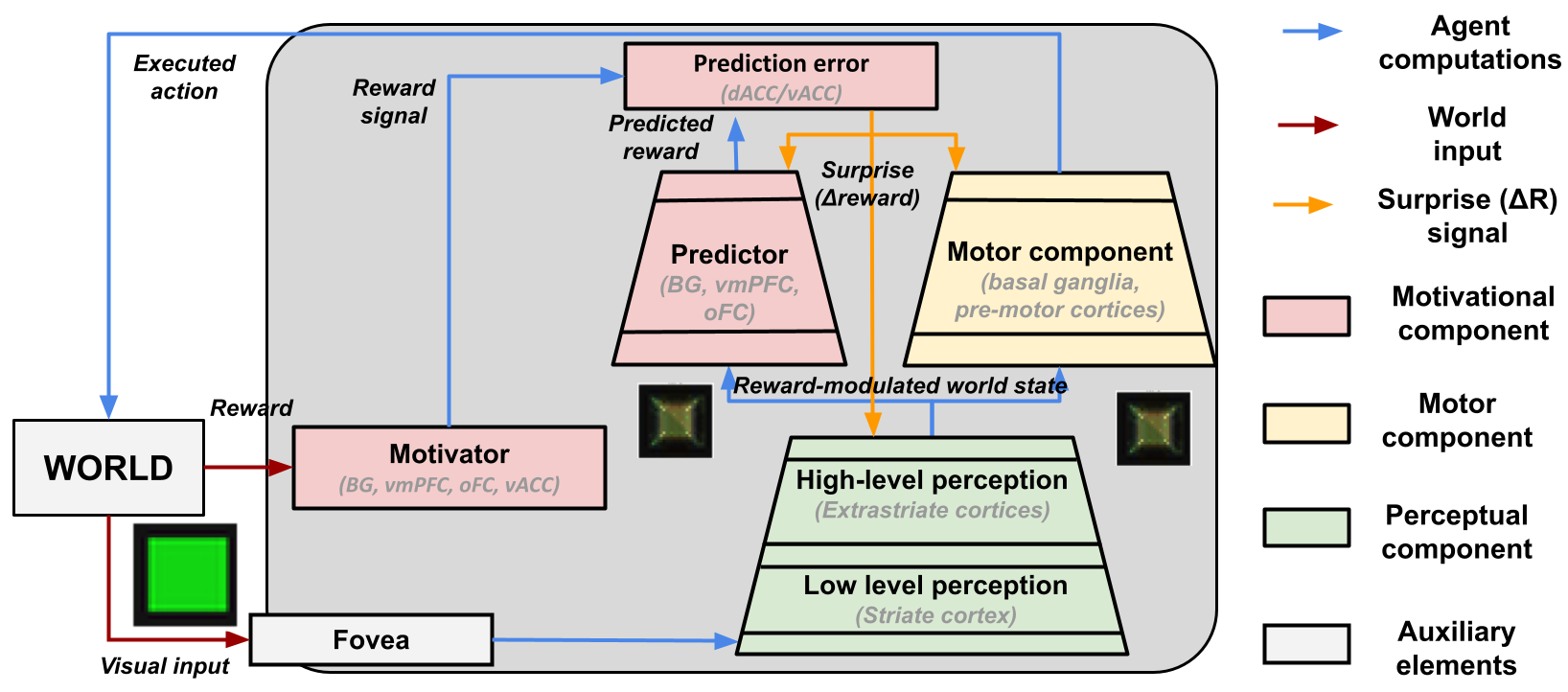}
	 \caption{Schema of the model components and functions, the flows of information between the components, and the learning signals.}
	 \label{Figure:FunctionalSchema}
\end{figure*}

Figure \ref{Figure:Brain_learning} summarises at a high level the elements of the hypothesis proposed here that are captured by the computational model.
This in particular encompasses intermediate layers corresponding to extra-striate cortices that host mixed UL and RL processes. 
Figure~\ref{Figure:FunctionalSchema} shows the architecture of the model.
We now illustrate the components of the architecture and their biological underpinning.

% Perceptual component
\paragraph{Perceptual component}
This component is based on a neural network that processes visual inputs by performing information abstraction and mimics the brain visual cortical system.
In particular, the component executes a hierarchical information processing \citep{FellemanEssen1991Distributedhierarchicalprocessing, BaldassarreCaligioreMannella2013Thehierarchicalorganisationofcorticalandbasalgangliasystemsacomputationallyinformedreviewandintegratedhypothesis} from the low-level retinotopic features in the striate cortex (V1), to the extraction of high-level image features (colour, shape, size) in the extrastriate cortices \citep{deyoe1996mapping, KonenKastner2008Twohierarchicallyorganizedneuralsystemsforobjectinformationinhumanvisualcortex}.

%Our model is bio-plausible
Differently from the biologically implausible gradient-descent methods, the network learns through a bio-plausible mechanism \citep{illing2019biologically}.
In particular, the learning processes used in the model update each connection weight (synapse) on the basis of locally available information related to the pre-synaptic and post-synaptic units.
Another biologically plausible feature of the model, at the core of the novelty of the hypothesis presented here, is that the top layer of this component is trained during the task through a mechanism that integrates associative and reward-based RL (Figure \ref{Figure:Brain_learning}).

The bottom layer of the component, which mimics early visual cortices, is instead trained before the task execution to reflect the learning of these areas during early development \citep{siu2018development}.
Critical for our hypothesis, this architecture captures the essence of the effects of dopamine reward signals onto extra-striate cortices and the lack of it in striate cortices \citep{ WilliamsGoldmanRakic1993CharacterizationoftheDopaminergicInnervationofthePrimateFrontalCortexUsingaDopamineSpecificAntibody,jacob2018monoaminergic, impieri2019receptor, niu2020receptor,froudist2020dopamine}.
Finally, the model relies on distributed representations, for which information on each content (e.g., a percept) is encoded by many units of the layer, and each unit takes part in the representations of different contents.
This encoding is more bio-plausible than localistic representations (`grandmother-cells'; \citealp{McClellandPDPResearchGroup1986ParallelDistributedProcessingExplorationsintheMicrostructureofCognition,quiroga2008sparse}).

% Motor component
\paragraph{Motor component} 
This component is supported by a neural network that, on the basis of the perceptual component activation, produces an `action' affecting the world.
The network is trained through a trial-and-error learning algorithm using a reward signal, mimicking the interactions of basal ganglia with motor cortices during the learning of actions \citep{kim2017reward, seger2008basal}. 

% Motivator component 
\paragraph{Motivational component}
%
% This component is formed by the typical modules implementing the functions of the \textit{critic} of the actor-critic model \citep{SuttonBarto2018ReinforcementLearningAnIntroduction}; these components might correspond to different brain sub-systems. 
%
This component is formed by three sub-modules that emulate the motivational functions supported by different brain sub-systems. 

% \textit{reward-signal}

First, a \textit{motivator} sub-module produces a reward signal on the basis of the perceived outcome following action performance.
Here the outcome is received from the environment and informs the system on the `correctness' of the performed action (see below).
This action-outcome might correspond to an `extrinsic reward', for example to the receipt of food or other rewarding resources; this is suitably processed by the system sensors and motivator component to produce a reward signal guiding the system learning processes.
%FEDE: sotto facciamo una lunga rev dell'intrisic motivation e ci sta. Pero' non la introdurrei con un semplice "in other conditions" perche' crea l'aspettativa nel lettore che siano condizioni del modello. Forse metterei "Alternatively". Comunque specificherei che non lo trattiamo qui. 
In other conditions \citep{BaldassarreMirolli2013,Baldassarre2011WhatAreIntrinsicMotivationsABiologicalPerspective} 
the reward signal might be produced by intrinsic motivation processes, for example related to the novelty or surprise of the experienced stimuli \citep{BartoMirolliBaldassarre2013Noveltyorsurprise} or to the acquisition of competence during the accomplishment of a desired goal \citep{WHITE1959,SantucciBaldassarreMirolli2016GRAILaGoalDiscoveringRoboticArchitectureforIntrinsicallyMotivatedLearning}. 
%For example in a peg-in-a-hole task a child might self-generate a reward signal based on the distance of the object position/orientation from the target hole where to insert the object \citep{taffoni2019motor}.
%
In the brain, structures such as the hypothalamus and the pedunculopontine nucleus, and the ventromedial, orbital, and anterior-cingulate cortices, support extrinsic rewards \citep{Panksepp1998AffectiveNeurosciencetheFoundationsofHumanandAnimalEmotions,MirolliMannellaBaldassarre2010Therolesoftheamygdalaintheaffectiveregulationofbodybrainandbehaviour}, while other structures, such as the superior colliculus, hippocampus, and the dorsolateral prefrontal cortex, support the computation of intrinsic reward signals \citep{LismanGrace2005TheHippocampalVTALoopControllingtheEntryofInformationintoLongTermMemory,RibasFernandesSolwayDiukMcGuireBartoNivBotvinick2011Aneuralsignatureofhierarchicalreinforcementlearning,Baldassarre2011WhatAreIntrinsicMotivationsABiologicalPerspective}.

Second, a \textit{predictor} sub-module, based on a multi-layer neural network, uses the high-level perceptual representations encoding the current perceived state, received from the top layer of the perceptual component, to predict the rewards that can be attained from it.
This module functionally mimics the brain basal-ganglia striosomes \citep{HoukDavidsBeiser1995ModelsofInformationProcessingintheBasalGanglia}.

Last, a \textit{prediction error} sub-module integrates the obtained and predicted rewards and produces a learning signal (`surprise'). This signal influences the learning of the predictor, of the motor component and, most importantly, of the perceptual component. 
In the brain, this signal is represented by the phasic dopamine bursts reaching various target areas \citep{Schultz2002GettingFormalwithDopamineandReward}, as also modelled by the actor-critic RL architecture \citep{Barto1995AdaptiveCriticsandtheBasalGanglia}.

%
% Auxiliary components
%\paragraph{Auxiliary elements} 
%These elements allow the model to process an external `realistic' input and to face task demands. 
%%
%%Retina
%The first function is supported by a simplified retina that is able to detect raw information of external input image encoded as RGB pixels. 
%%
%%Enviroment
%The second component acts as an abstract environment that:
%(a) provides input patterns;
%(b) provide a reward informing the system on the correctness of the action, based on the visual features of the input image (i.e. colour/shape/size).
%Abstracting a physical environment, this function allows the single-trial execution and the related computation of agent performances by an evaluator, i.e. the distance between an executed action and the correct action suggested by the environment.
%
%Elements of embodiment (soft)
%Despite simplification, these components give the model some elements of embodiment \citep{da2018defining}.

\subsection{Computational implementation and learning algorithms of the model}
\label{Section:computational_details}

\begin{figure*}[htb!]
    \centering
	\includegraphics[width=\textwidth]{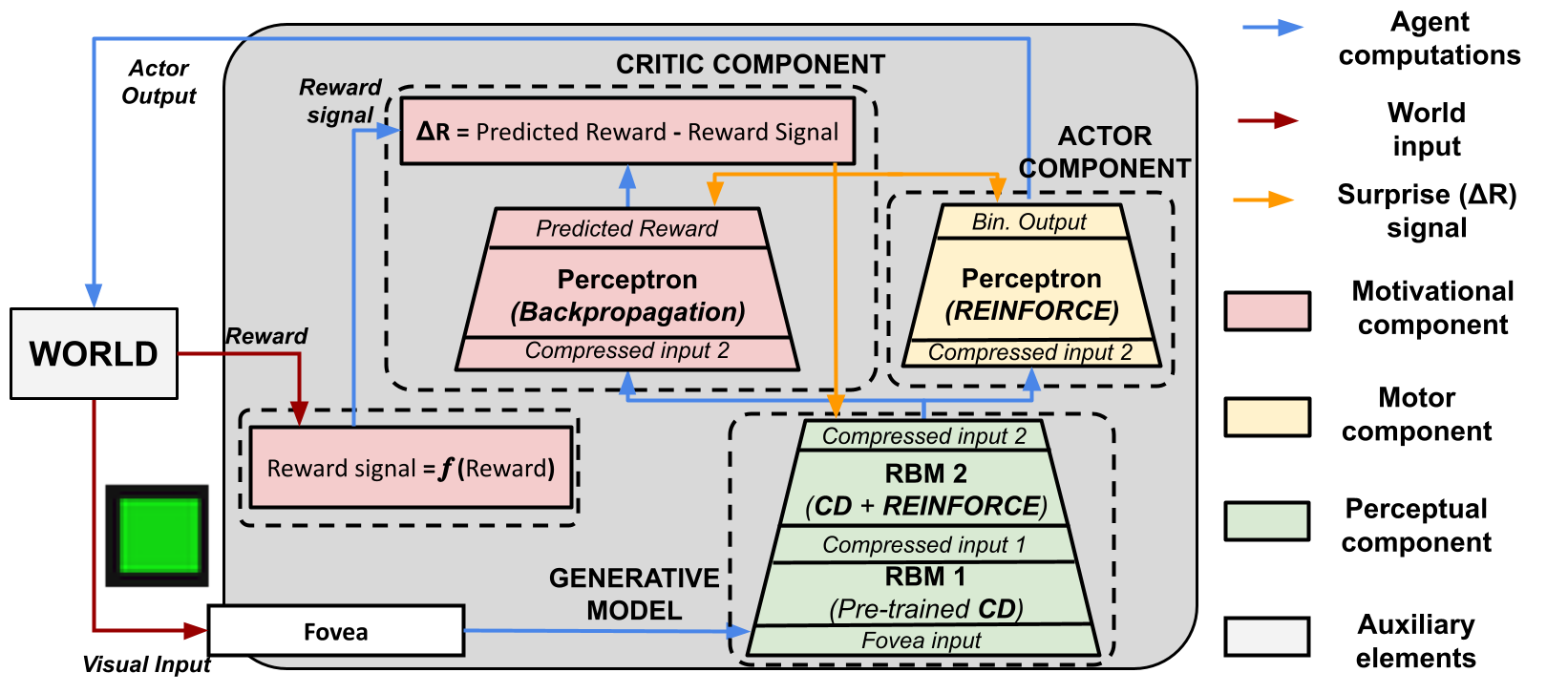}
	 \caption{A computational schema of the model components and their training algorithms, the flows of information between the components, and the learning signals.
	 }
	 \label{Figure:ComputationalSchema}
\end{figure*}

The system proposed here (Figure~\ref{Figure:ComputationalSchema}) is formed by a generative model integrated into an actor-critic architecture \citep{sutton1998reinforcement}, both modified to study the role of reward in perceptual representation learning.
Further details regarding the system parameters (e.g., the number of units of each layer, the learning rates, the training epochs, etc.) are reported in Table S1 of the Supplementary Materials.
The code of the system will be made publicly available online in GitHub in the case of publication.
%Federico: perche' non mettiamo subito il link?

% Perceptual component - DBN
\paragraph{Perceptual component} 

This component is a generative \textit{Deep Belief Network} (DBN; \citealp{hinton2006fast,le2008representational}) composed of two stacked \textit{Restricted Boltzmann Machines} (RBM; \citealp{Hinton2012}).
Each RBM is composed of an input layer (`visible layer') and a second layer (`hidden layer') formed by Bernoulli-logistic stochastic units where each unit $j$ has an activation $h_j\in\{0,1\}$:
\begin{align}
    h_j &= \begin{cases}1 \ if\  \nu \ge \sigma(p_j) \\0\  if\  \nu < \sigma(p_j) \end{cases}
    \\
    \sigma(p_j) &= \frac{1}{1+e^{-p_j}} \nonumber
    \\
    p_j &= \sum_i(w_{ji} \cdot v_i) \nonumber
    \label{eq_Bernoulli}
\end{align}
where 
$\sigma(x)$ is the sigmoid function, 
$p_j$ is the activation potential of the unit $h_j$, 
$\nu$ is a random number uniformly drawn from $(0,1)$ for each unit, and 
$w_{ji}$ is the connection weight between the visible unit $v_i$ and $h_j$.
The RBM is capable of reconstructing the input by following an inverse activation from the hidden layer to the input layer.

% re-consider as a second stochastic layer (using the same mechanisms illustrated  for the hidden layer).

The DBN consists of a stack of RBMs---two in the model---where each RBM receives as input the activation of the hidden latent layer of the previous RBM.
The model is trained layer-wise, starting from the RBM which receives inputs from the environment and towards the inner layers.  
On this basis, the DBN executes an incremental dimensionality reduction of the input, as higher layers further compress the representations received from the lower/previous RBM \citep{hinton2006reducing}. 
In the model, the first RBM directly receives the input images and it is trained to encode them  `offline' before the task.
This training uses the \textit{Contrastive Divergence}, an unsupervised-learning algorithm that computes each connection weight update $\Delta\it{w_{i j}}$ 
% based on four  visible-hidden-visible-hidden succeeding activations 
 as follows:
\begin{equation}
    \Delta\it{w_{i j}}=\epsilon(\langle \it{v_i \cdot h_j} \rangle_{\it{data}} - \langle \it{v_i \cdot h_j} \rangle_{\it{model}})
    \label{eq_weight_update_CD}
\end{equation}
where
$\epsilon$ is the learning rate,
$\langle \it{v_i \cdot h_j} \rangle_{\it{data}}$ is the product between the initial input (initial visible activation) and the consequent hidden activation averaged over all data points,
$\langle \it{v_i \cdot h_j} \rangle_{\it{model}}$ is the product between the reconstructed visible activation and a second activation of the hidden layer following it averaged over all data points.
The third and fourth activations are denoted with `model' as they tend to more closely reflect the spontaneous input-independent activations
% firing 
 of the RBM.

The second RBM of the model is trained `online' during the task performance based on the novel algorithm proposed here.
The algorithm integrates \textit{Contrastive Divergence} (Eq.~\ref{eq_weight_update_CD}) with the \textit{REINFORCE} algorithm described in the next session (Eq.~\ref{REINFORCE_with_Bernoulli_logistic}) as follows:
\begin{equation}
\begin{split}
\Delta\it{w_{i j}} = & \ \lambda \ (\epsilon \ (\langle \it{v_i \cdot h_j} \rangle_{\it{data}} - \langle \it{v_i \cdot h_j} \rangle_{\it{model}})) \ + 
\\
& \ (1 - \lambda) \ (\alpha \ (r - \bar{r}) (y_{\it{j}} - p_{\it{j}}) x_{\it{i}})
\end{split}
\end{equation}
where 
$\lambda$ is the contribution of Contrastive Divergence to the update of weights, and
$(1 - \lambda)$ the contribution of REINFORCE.
Crucial for this work, $\lambda$ mixes the contribution of UL and RL processes to the weight update, in particular a high value of it implies a dominance of UL whereas a low value of it implies a dominance of RL.
In the simulations, we tested five values of the parameter: $\lambda \in \{1, 0.1, 0.01, 0.001, 0\}$.

% Motor component - RL perceptron (WIlliams)
\paragraph{Motor component} 

This component is a single-layer neural network trained with the RL algorithm REINFORCE \citep{williams1992simple}.
The input of the network is the activation of the last layer of the perceptual component.
The network output layer is composed of Bernoulli-logistic units as for the perceptual component.
The algorithm computes the update $\Delta\it{w_{ji}}$ of each connection weight linking the input unit $i$ and the output unit $j$ of the component as follows:
\begin{equation}
    \Delta\it{w_{ji}} = \alpha (r - \bar{r}) (y_{\it{j}} - \sigma(p_{\it{j}})) x_{\it{i}}
    \label{REINFORCE_with_Bernoulli_logistic}
\end{equation}
where
$\alpha$ is the learning rate,
$r$ is the reward signal received from the motivator,
$\bar{r}$ is the reward signal expected by the predictor,
$x_{\it{i}}$ is the input of the network (from the outer second hidden layer of the DBN),
$\sigma(p_{\it{j}})$ is the sigmoidal activation potential of the unit encoding its probability of firing, and
$y_{\it{j}}$ is the unit binary activation.

% Motivational component
\paragraph{Motivational component}

This component implements the functions of the \textit{critic} component of an \textit{actor-critic} architecture \citep{sutton1998reinforcement} based on three sub-modules introduced in the previous section. 

% Motivator
The \textit{motivator} module computes the reward signal by scaling the reward perceived from the external environment into a standard value, the \textit{reward signal} $r\in(0,1)$:
\begin{equation}
r =  f(Reward)
\end{equation}
where
$Reward$ is the reward perceived from the environment and $f(.)$ is a linear scaling function ensuring that the reward signal ranges between $0$, corresponding to a wrong action, to $1$, corresponding to an optimal action.
This reward signal represents the pivotal guidance of the RL processes driving the acquisition of not only the actions but also the DBN second hidden layer. 
As discussed above, in other cases the motivator might compute more complex reward signals based on more sophisticated types of extrinsic and/or intrinsic motivation mechanisms.

%Preditor
The \textit{predictor} module is a multi-layer perceptron composed of an input layer (DBN second hidden layer), a hidden layer, and an output linear unit predicting the expected reward signal $\bar{r}$.
The perceptron is trained with a standard gradient descendent method \citep{McClellandPDPResearchGroup1986ParallelDistributedProcessingExplorationsintheMicrostructureofCognition,amari1993backpropagation}
%\citep{WidrowHoff1960AdaptiveSwitchingCircuits}
using a learning rate $\alpha$ and the error $e$ computed by the prediction-error component.

%Prediction-error component
The \textit{prediction error} module is a function that computes the reward prediction error (surprise) $e$ as follows:
\begin{equation}
    e = r - \bar{r}
    \label{Supervised_learning_rule}
\end{equation}
where
$r$ is the reward signal from the motivator, and
$\bar{r}$ is the expected reward signal produced by the evaluator.
This error is used to train the predictor itself, the motor component, and the perceptual component. 

% Auxiliary elements
\paragraph{Auxiliary elements}

% Input Dataset
The input dataset is formed by RGB images with a black background and a polygon at the centre (Figure \ref{Figure:Task}).
The polygon is characterised by a unique combination of specific attributes chosen from three visual categories: colour, form and size.
There are four attributes for each category: 
red, green, blue, yellow (colour);  square, circle, triangle, bar (form);
large, medium-large, medium-small, small (size).
These attributes generate $4^3=64$ combinations forming the images used in the test.

% Retina
The retina component is implemented as a $28\times28\times3$ matrix containing the RGB visual input.
The matrix is unrolled into a vector of $2,352$ elements that represents the input of the perceptual component. 

%Environment
The environment is implemented as a function that provides an image to the model at each trial.
In one trial the model perceives and processes one input image and undergoes a cycle of the aforementioned learning processes based on the reward received from the environment after the action performance.
Here the environment computes the reward $r'$ simply on the basis of the Euclidean distance between the model action and an `optimal action':
\begin{equation}
Reward =  \|\mathbf{y}^* - \mathbf{y} \|_{1}
\end{equation}
where
$\mathbf{y}^*$ is the optimal action binary vector that the model should produce for the current input,
$\mathbf{y}$ is the model binary action, and
$\|.\|_{1}$ is the L1 norm of the vectors difference.
The optimal actions are four binary random vectors that the model should produce in correspondence to the items of the four input categories of the given task.

\section{Results}
%Argomenti centrali dei risultati
%Sistema misto e' il migliore.
%Il sistema tutto unsupervised va male perche':
%- Meno efficace (a steady state): ma solo se risorse scarse computazionali
%Il sistema tutto RL va male perche':
%- Meno efficiente
%- Si infila in minimi locali 

We tested the model with different tasks each involving one out of three sorting rules based on the three categories, for example, a task required sorting the cards by colour and another one by shape. 
The model was tested with five different levels of UL/RL contribution ($\lambda$ parameter, see Section \ref{Section:computational_details}) and two levels of internal resources, in particular respectively $10$ and $50$ units in the second DBN hidden layer.
Note that $50$ units were sufficient to allow the system to fully encode the image features, as shown by a preliminary test indicating a close-to-null reconstruction error.

We varied the parameters of these environmental and model conditions with a random grid search based on $1016$ simulations. The simulations were run in the \textit{Neuroscience Gateway platform} \citep[NSG,][]{sivagnanam2013introducing}.
The different values of the critical parameter $\lambda$ gave rise to five conditions labelled as follows:
Level 0 (L0): no RL (i.e., only UL);
Level 1 (L1): low RL;
Level 2 (L2): moderate RL;
Level 3 (L3): high RL;
Level 4 (L4): extreme RL (no UL). 
The simulations using different amounts of internal resources allowed us to investigate how the available computational resources affect the results related to the UL/RL mix.

The presentation of results is organised in three parts. 
%
% performance analysis
The first part investigates the effects on the performance of the different contributions of UL/RL. 
%
% Internal representation analyses
The second part investigates the nature of the perceptual representations acquired through a different UL/RL learning mix. 
%
% Sensory information stored
Finally, the third part presents a graphical reconstruction of the original input patterns produced by the generative perceptual component of the model and the related reconstruction errors.

\paragraph{Performances analysis} 

\begin{figure*}
    \centering
    \textbf{Learning curves of models} \par\medskip
	\includegraphics[width=0.8\textwidth, height = 18EM]{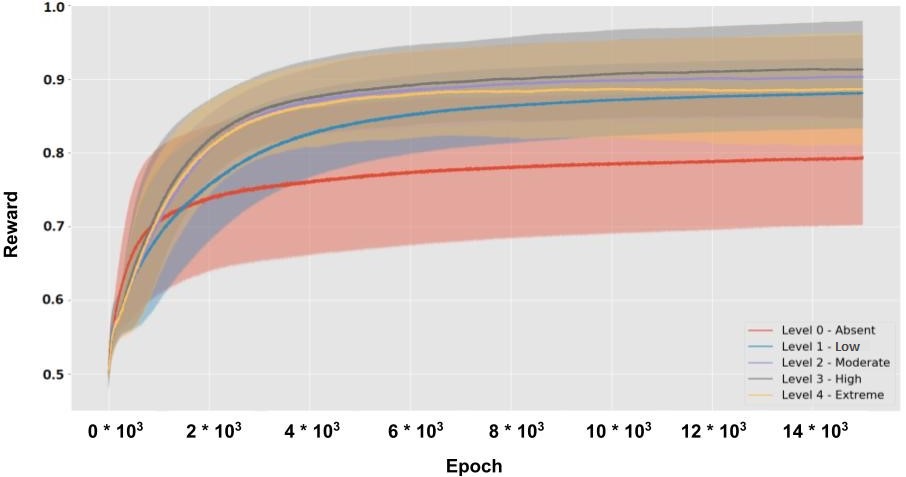}
	 \caption{Reward per epoch of the five models involving different UL/RL levels, averaged over the models using a given level.
	 Shaded areas represent the curves standard deviations.}
	 \label{Figure:Training_Curves}
\end{figure*}

%Training curves
%fede: dobbiamo spostare le figure tipo la 5 qui perche' c'e' inconsistenza tra la pag dove sono citate per la prima volta e la loro posizione
Figure \ref{Figure:Training_Curves} shows the training curves of the models trained with different RL contributions in 15,000 epochs.
The L0 models, using only UL, learn faster during the first 1,000 epochs but exhibits the worst final performance.
Figures S1, S2 and S3 in Supplementary Materials show that this effect is also present in subsets of all simulations.
Instead, the highest final performance is achieved by the L3 models where UL and RL are better balanced.

\begin{figure*}[htb!]
    \centering
    \textbf{Performance of models} \par\medskip
	\includegraphics[width=0.9\textwidth, height = 18EM]{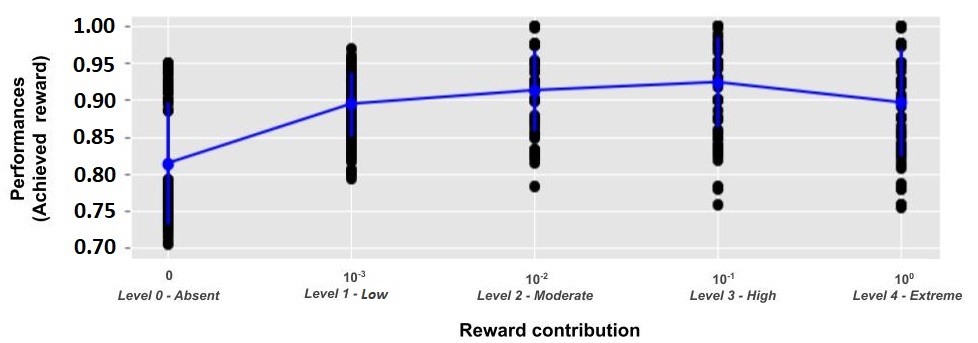}
	 \caption{Performances (maximum reward obtained at the end of training) of models featuring different levels of RL contribution.}
	 \label{Figure:MAX_R}
\end{figure*}

%Final rewards: correlations
Figure \ref{Figure:MAX_R} shows the final performance of the models, namely the maximum reward they achieved.
%at the end of training. 
%
A correlation analysis shows the presence of a linear relation between such performance and the level of the RL, but this is not very high thus indicating the relevance of the inverted U shape of the curve visible from the figure ($r = 0.5$, $p < 0.001$).
%
% Anova
A one-way ANOVA confirms the presence of a statistical difference between the final performance of the five groups ($F = 47.51$, $p < 0.001$).
%
%Post hocs
Post hoc tests (Table \ref{Table:Post_Hoc}) confirm that the performances of models with an absent RL contribution (L0) are statistically different with respect to each of the other models ($0.81 \pm 0.08$, $p < 0.001$). 
The L3 models show a higher performance compared to the
L0 models ($0.92 \pm 0.06$ vs. $0.81 \pm 0.08$, $p < 0.001$), the 
L1 models ($0.92 \pm 0.06$ vs. $0.89 \pm 0.04$, $p < 0.001$), and the 
L4 models ($0.92 \pm 0.06$ vs. $0.90 \pm 0.07$, $p < 0.05$).
The L2 and L3 models do not show a significant difference ($0.92 \pm 0.06$ vs. $0.91 \pm 0.05$).

\begin{table*}[htb!]
\centering
\begin{tabular}{|c | c | c | c | c | c|}
\hline
 & \textbf{Absent (L0)} & \textbf{Low (L1)} & \textbf{Moderate (L2)} & \textbf{High (L3)} & \textbf{Extreme (L4)} \\
    \hline
   \textbf{Absent (L0)}        & //  &  // & // & // & // \\
    \hline
    \textbf{Low (L1)}        & $p < 0.001$ & //  & // & // & // \\
    \hline
    \textbf{Moderate (L2)}       & $p < 0.001$  & $p > 0.05$ (NS)  & // & // & // \\
    \hline
    \textbf{High (L3)}       & $p < 0.001$  & $p < 0.001$  & $p > 0.05$ (NS) & //  & // \\
    \hline
    \textbf{Extreme (L4)}        & $p < 0.001$  & $p > 0.05$ (NS)  & $p > 0.05$ (NS) & $p < 0.05$ & // \\
    \hline
\end{tabular}
\caption{Post-hoc comparisons (t-test with Bonferroni correction) between the performance of models with different levels of RL contribution.
`NS' indicates `non statistically significant'.}
\label{Table:Post_Hoc}
\end{table*}

% \begin{table*}[htb!]
% \centering
% \begin{tabular}{|c | c | c | c | c | c|}
% \hline
%  & \textbf{Absent (L0)} & \textbf{Low (L1)} & \textbf{Moderate (L2)} & \textbf{High (L3)} & \textbf{Extreme (L4)} \\
%     \hline
%   \textbf{Absent (L0)}        & //  & $p < 0.001$ & $p < 0.001$ & $p < 0.001$ & $p < 0.001$ \\
%     \hline
%     \textbf{Low (L1)}        & $p < 0.001$ & //  & $p > 0.05$ (NS) & $p < 0.001$ & $p > 0.05$ (NS) \\
%     \hline
%     \textbf{Moderate (L2)}       & $p < 0.001$  & $p > 0.05$ (NS)  & // & $p > 0.05$ (NS) & $p > 0.05$ (NS) \\
%     \hline
%     \textbf{High (L3)}       & $p < 0.001$  & $p < 0.001$  & $p > 0.05$ (NS) & //  & $p < 0.05$ \\
%     \hline
%     \textbf{Extreme (L4)}        & $p < 0.001$  & $p > 0.05$ (NS)  & $p > 0.05$ (NS) & $p < 0.05$ & // \\
%     \hline
% \end{tabular}
% \caption{Post-hoc comparisons (t-test with Bonferroni correction) between the performance of models with different levels of RL contribution. `NS' indicates `non statistically significant'.}
% \label{Table:Post_Hoc}
% \end{table*}

To further investigate the relationship between the performance of the models and the different levels of RL contribution, we grouped the results of the simulations on the basis of the computational resources or the sorting rule.
Here we present a summary of the results while Section S2.1 in the Supplemental Materials reports the posthoc tests.

% resources-specific differences
Table \ref{Table:Resources_Categories_conditions} shows that the increase of computational resources available for the representations tends to %cause a ceiling effect by lowering the relationship between the RL contribution and the performance level (Pearson correlation: $\rho=0.5$, $p < 0.001$ vs. $\rho=0.1$, $p < 0.05$).
lower the amount of RL contribution leading to the highest performance.
Indeed, a one-way ANOVA shows a statistical difference between the models ($F = 3.85$, $p < 0.001$) and the post-hoc tests show that the L2 model leads to the best result ($0.95 \pm 0.05$).

% category-specific learning
The table also highlights differences between the simulations using different sorting rules (colour, shape, size). 
%
%Colour
The simulations with the \textit{colour sorting rule} show 
%strong ceiling effect, causing 
flattened reward values with respect to the different RL contribution.
In the case of low computational resources the model does not show statistically significant differences ($F = 0.88$, $p > 0.05$).
A difference emerges in the case of high computational resources ($F = 19.8$, $p < 0.001$)
where the L2 models, having a balanced UL/RL mix, show the best final performance ($0.98 \pm 0.02$). 

%Shape
The simulations with the \textit{shape sorting rule} show statistical differences with both low computational resources ($F = 120.9 $, $p < 0.001$) and high computational resources ($F = 20.4 $, $p < 0.001$).
In both cases, the models using a mixed level of UL and RL prevail:
the extreme cases of the L0 models (only UL), and L4 models (only RL) have lower performances with respect to the L1, L2 and L3 models having a more balanced UL/RL mix.

% size
Finally, the simulations with the \textit{size sorting rule} show statistical differences with low computational resources ($F = 43.4 $, $p < 0.001$) but not with `high computational resources' ($F = 1.12 $, $p > 0.05$).
In the first case, the L0 models have the lowest performance.

% \begin{table*}[htb!]
% \centering
% \begin{tabular}{|c | c | c | c | c | c|}
% \hline
%  & \textbf{Absent} & \textbf{Low} & \textbf{Moderate} & \textbf{High} & \textbf{Extreme} \\
%     \hline
%   \textbf{Low resources}        & $0.81 \pm 0.08$  & $0.89 \pm 0.04$ & $0.91 \pm 0.05$ & $0.92 \pm 0.06$ & $0.90 \pm 0.07$ \\
%     \hline
%     \textbf{High resources}        & $0.92 \pm 0.03$  & $0.93 \pm 0.04$ & $0.95 \pm 0.05$ & $0.93 \pm 0.06$ & $0.93 \pm 0.05$ \\
%     \hline
% \end{tabular}
% \caption{The table shows the performances of models in case of `low resources' condition (less neurons in the last layer of perceptual component) and `high resources' condition (more neurons in the last layer of perceptual component).}
% \label{Table:Resources_conditions}
% \end{table*}

\begin{table*}[htb!]
\centering
\begin{tabular}{|c | c | c | c | c | c|}
\hline
  & \textbf{Absent} & \textbf{Low} & \textbf{Moderate} & \textbf{High} & \textbf{Extreme} \\
    \hline
    \textbf{Low Resources} & $0.81 \pm 0.08$  & $0.89 \pm 0.04$ & $0.91 \pm 0.05$ & $\mathbf{0.92} \pm 0.06$ & $0.90 \pm 0.07$\\
    \hline
   \textbf{Colour}        & $0.92 \pm 0.02$  & $\mathbf{0.92} \pm 0.02$ & $0.91 \pm 0.04$ & $0.91 \pm 0.07$ & $0.90 \pm 0.08$ \\
    \hline
    \textbf{Shape}        & $0.75 \pm 0.02$  & $0.89 \pm 0.04$ & $0.94 \pm 0.04$ & $\mathbf{0.95} \pm 0.04$ & $0.93 \pm 0.06$ \\
    \hline
    \textbf{Size}        & $0.76 \pm 0.02$  & $0.88 \pm 0.05$ & $0.89 \pm 0.06$ & $\mathbf{0.90} \pm 0.06$ & $0.86 \pm 0.07$ \\
    \hline
    \textbf{High Resources} & $0.92 \pm 0.03$  & $0.93 \pm 0.04$ & $\mathbf{0.95} \pm 0.05$ & $0.93 \pm 0.06$ & $0.93 \pm 0.05$  \\
    \hline
   \textbf{Colour}        & $0.94 \pm 0.01$  & $0.94 \pm 0.01$ & $\mathbf{0.98} \pm 0.02$ & $0.95 \pm 0.03$ & $0.96 \pm 0.02$ \\
    \hline
    \textbf{Shape}        & $0.93 \pm 0.02$  & $0.97 \pm 0.02$ & $\mathbf{0.97} \pm 0.02$ & $0.96 \pm 0.02$ & $0.94 \pm 0.02$ \\
    \hline
    \textbf{Size}        & $0.88 \pm 0.02$  & $0.88 \pm 0.03$ & $\mathbf{0.90} \pm 0.05$ & $0.88 \pm 0.07$ & $0.88 \pm 0.07$ \\
    \hline
\end{tabular}
\caption{Performance of models with different RL contributions in correspondence to two different amounts of computational resources (number of neurons in the second hidden layer of the DBN) and three different sorting rules (colour, shape, size).
Figures in bold highlight the highest value per each condition (along the raws).}
\label{Table:Resources_Categories_conditions}
\end{table*}

\paragraph{Analysis of internal representations}

To investigate the nature of the perceptual representations acquired by the models, we show the results of some example simulations in the cases of different sorting rules and different levels of the RL.
Other simulations lead to qualitatively similar results.

To visualise the representations we used a Principal Component Analysis (PCA), allowing a dimensionality reduction, and a K-means algorithm, supporting clustering.
In particular, we extracted the first two principal components of the input patterns reconstructed by the model into the visible layer in correspondence to the original 64 input patterns.
The reconstructed images were obtained by spreading the activity from the visible layer of the DBN, activated with an image, to its first and second hidden layer, and then back towards the visible layer.
% (as it is commonly done to have stable data with few computations, we directly spread the sigmoidal activity of the units without the stochastic binary extraction).
We analysed the reconstructed visual representations, rather than the hidden representations,
to assess which features of the original visual images are retained by the internal representations. 
%ho capito cosa vuoi dire ma non si capisce. Non saprei come girarla mantenedo la chiarezza. Secondo me come l'ho proposta e' suff. Nel senso, il lettore capisce.
% as they allow the performance of the PCA and K-means clustering but at the same time they also allow the display of which features of the original visual images are retained by the internal representations.
The results of the PCA extraction of the two dimensions of the representations can be plotted in a 2D scatter plot to visualise the results of the following K-means algorithm.
The K-means algorithm was applied to 
PCA 2D codes  %mi suona strano 2D code. Se lo hai letto da qualche parte bene altrimenti mnetterei "the first 2 components of the PCA"
of the internal representations.
We set $K=4$, so the algorithm grouped the representations into four classes, as the number of the actions.
This made it possible to analyse how the model internally represents the different input images. 
Further details and results regarding these methods, as the cumulative variance explained by the PCA components and the silhouette scores of the K-mean algorithm, are reported in Section S2.2 of Supplemental Materials.

\begin{figure*}[htb!]
    \centering
    \textbf{Colour sorting category: reconstructed input} \par\medskip
	\includegraphics[width=\textwidth]{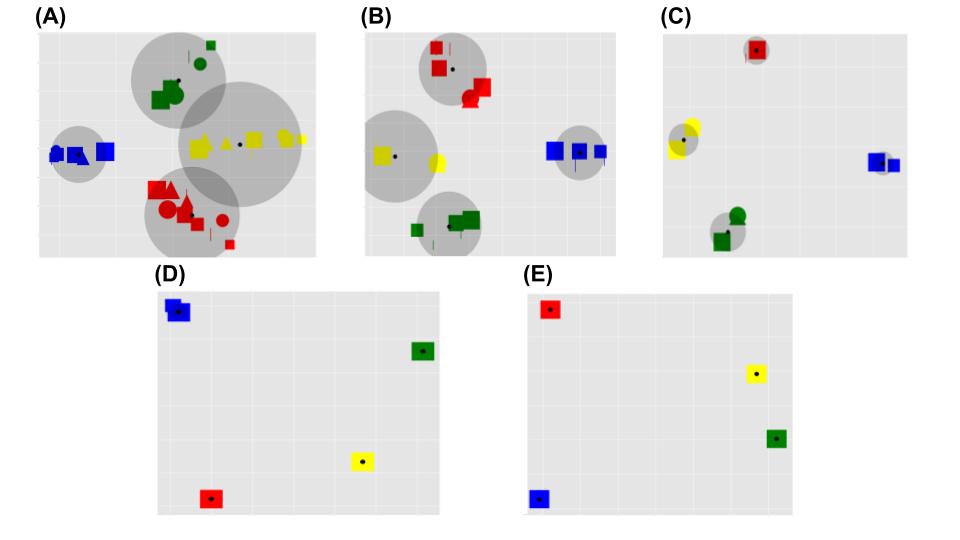}
	 \caption{Principal components of the reconstructed image representations in the case of the colour sorting rule and in correspondence to different levels of RL (shown in different graphs).
	 The dimensionality of the reconstructed image was reduced to two through a PCA  (x-axis: first component; y-axis: second component).
	 Within each graph, each reconstructed image is represented by a point marked by an icon that summarises the colour, shape, and size of the shape in the image (some icons are not visible as they overlap).
	 The centroids of the four clusters found by the K-means algorithm are marked with a black dot, while the maximum distance of the points of the cluster from its centroid is shown by a grey circle.
	 A: Level 0 (L0), absent RL (only UL); 
     B: Level 1 (L1), low RL;
     C: Level 2 (L2), moderate RL;
     D: Level 3 (L3), high RL;
     E: Level 4 (L4), extreme RL (no UL).
	 }
	 \label{Figure:Internal_Colour}
\end{figure*}

\begin{figure*}[htb!]
    \centering
    \textbf{Shape sorting category: reconstructed input} \par\medskip
	\includegraphics[width=\textwidth]{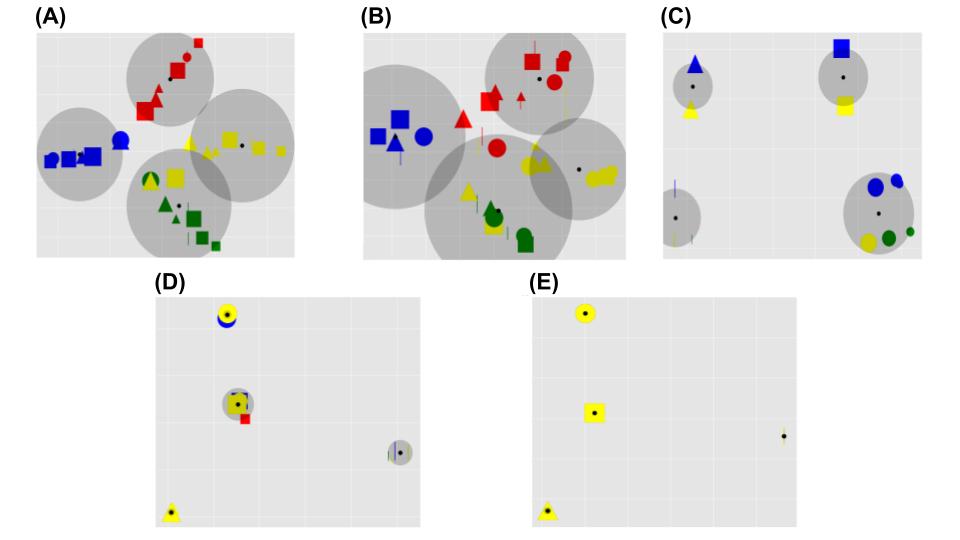}
	 \caption{Principal components of the reconstructed image representations in the case of the shape sorting rule and in correspondence to different levels of RL.
	 The plots are drawn as in Figure~\ref{Figure:Internal_Colour}.}
	 \label{Figure:Internal_Shape}
\end{figure*}

\begin{figure*}[htb!]
    \centering
    \textbf{Size sorting category: reconstructed input} \par\medskip
	\includegraphics[width=\textwidth]{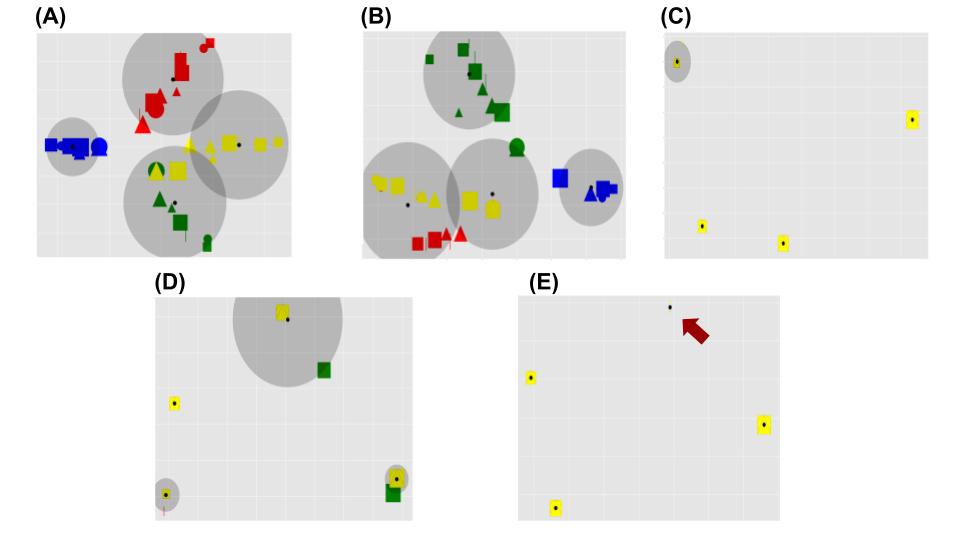}
	 \caption{Principal components of the reconstructed image representations in the case of the size sorting rule and in correspondence to different levels of RL.
	 The graphs are drawn as in Figure~\ref{Figure:Internal_Colour}.
	 The red arrow in graph E indicates the centroid of a cluster that contains only the small bars but not the other small shapes.
	 }
	 \label{Figure:Internal_Size}
\end{figure*}

% PCA plots
The results of the analyses (Figures~\ref{Figure:Internal_Colour}-\ref{Figure:Internal_Size}) highlight that the reward-based RL contribution strongly affects the internal representations as revealed by the reconstructed inputs.
For each sorting rule considered, models with a medium (L2) and high (L3) level of RL show the emergence of
category-based clusters, with their radius progressively diminishing with an increasing weight of the RL. Conversely, the L0 and L1 models do not show this effect in any task condition.
%
% Exceptions: colour also in absence of reward
The only exceptions to this are the models with an absent or low RL (L0 and L1) showing a clustering effect that does not depend on the task but only on the colour of the shapes.
This is due to the high distinctiveness of colours, largely activating different portions of the input units with respect to the other image features.

%Error in case of size and extreme RL
Figure \ref{Figure:Internal_Size}-E shows that the model with an extreme RL incurred a clustering error.
In particular, in this condition the model should group the images into four clusters (as in the conditions of Figure~\ref{Figure:Internal_Size}-C,D) whereas it tends to use only three clusters. 

\paragraph{Information stored by the model} 

To further investigate what type of information is stored by the model, we show the results of two additional analyses.
The first analysis examined the DBN reconstruction error (see Section~S2.3 in Supplementary Materials for further details),
while the second analysis qualitatively inspected the reconstructions of the input images.

\begin{figure*}[htb!]
    \centering
    \textbf{Information loss for different levels of RL} \par\medskip
	\includegraphics[width=0.9\textwidth, height = 18EM]{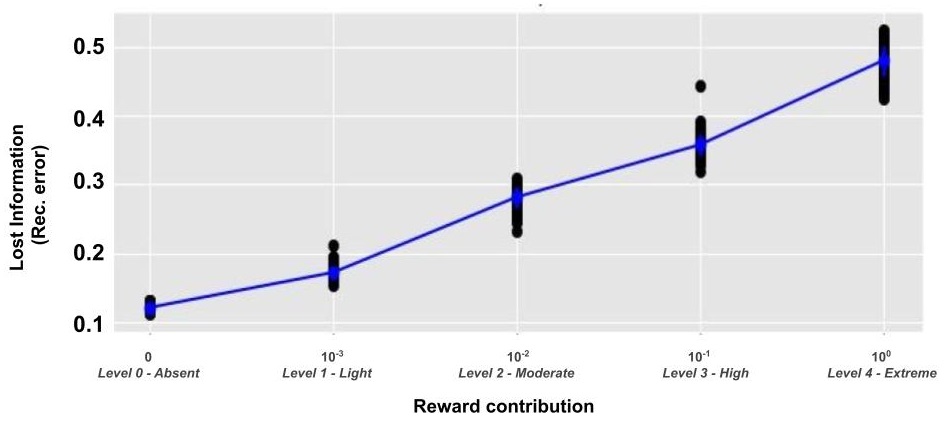}
	 \caption{Information loss (reconstruction error at the end of the training) of models with different levels of RL.}
	 \label{Figure:Min_REc_Error}
\end{figure*}

% correlation
Figure \ref{Figure:Min_REc_Error} shows the results of the first analysis and highlights the presence of a strong positive linear relationship between the level of RL and the reconstruction error ($ r = 0.68$, $p < 0.001$).  
%
% Anova
A one-way ANOVA confirmed the presence of a statistical difference between the five groups ($F > 100.0$, $p < 0.001$).
These results indicate that an increasing RL contribution causes a progressive loss of information on the input images.

\begin{figure}[htb!]
    \centering
    \textbf{Input reconstructions (sorting category: colour)} \par\medskip
	\includegraphics[width=0.5\textwidth]{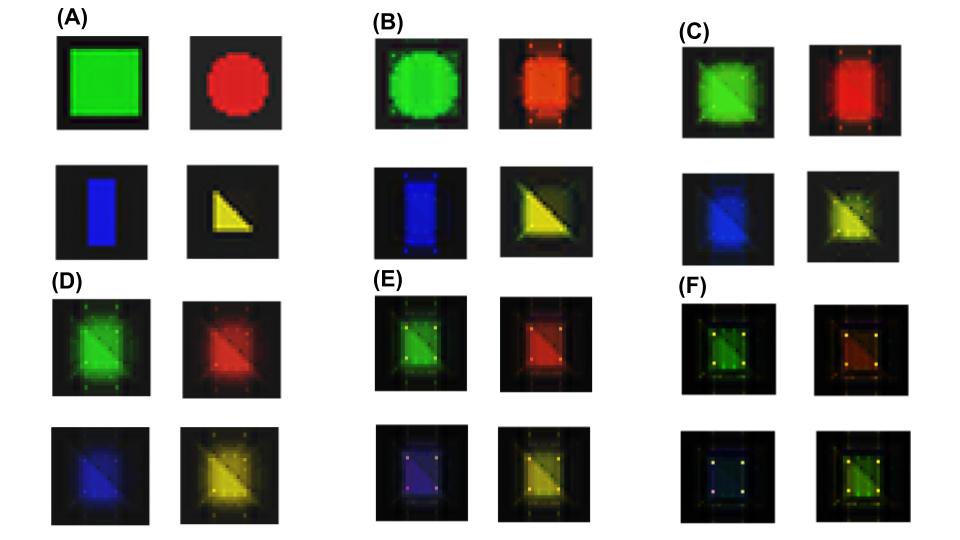}
	 \textbf{Input reconstructions (sorting category: shape)} \par\medskip
	\includegraphics[width=0.5\textwidth]{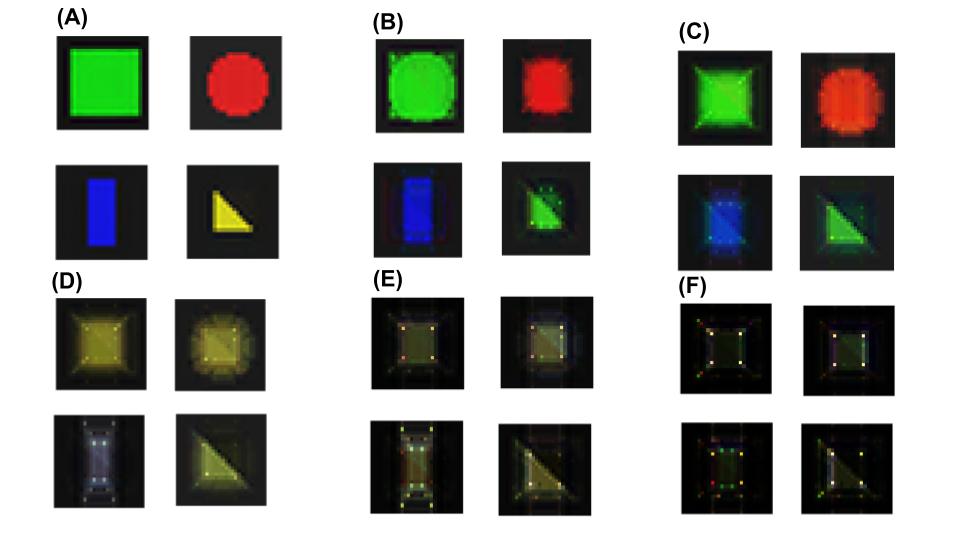}
	\textbf{Input reconstructions (sorting category: size)} \par\medskip
	\includegraphics[width=0.5\textwidth]{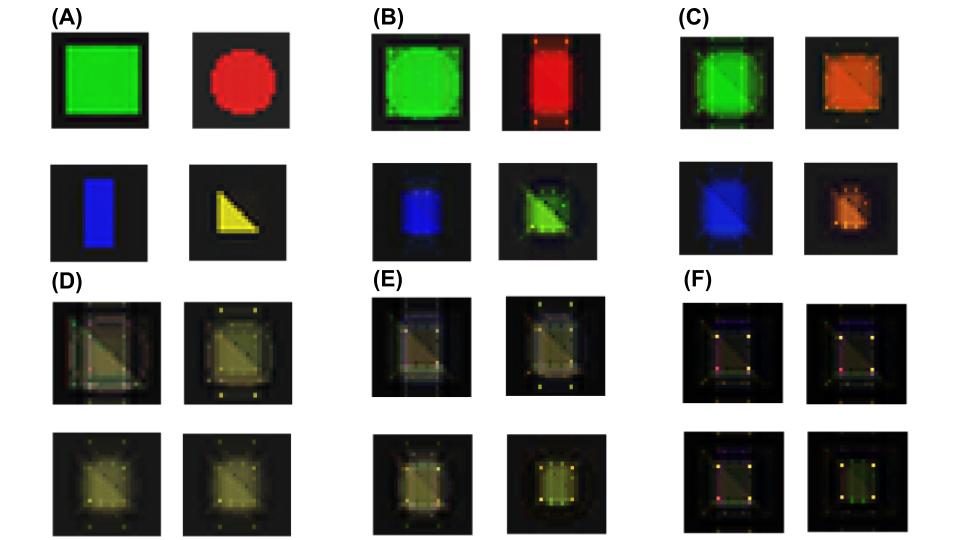}
	 \caption{Image reconstructions with different sorting rules and different levels of RL.
	 A: Original inputs;
 	 B: Level 0 (L0) - absent RL (only UL); 
     C: Level 1 (L1) - low RL;
     D: Level 2 (L2) - moderate RL;
     E: Level 3 (L3) - high RL;
     F: Level 4 (L4) - extreme RL (only RL).}
	 \label{Figure:Reconstructions}
\end{figure}

% Reconstruction
The results of the second analysis show the kind of information that the internal representations tend to retain, in particular if the system tends to store task-independent and/or task-related features.
In this respect, Figure~\ref{Figure:Reconstructions} highlights the emergence of shapeless coloured blobs in the case of the colour sorting rule, 
the emergence of colourless and sizeless prototypical shapes in case of shape sorting rule, and 
the emergence of colourless blobs with different sizes in the case of the size sorting rule.

\section{Discussion}

In this work, we propose a novel hypothesis on the broad nature of learning in the brain cortex.
A previous seminal view proposed that the cortex, basal ganglia, and cerebellum use the different learning mechanisms studied in machine learning, 
respectively %FEDE: respectively e' pericoloso perche' non si usa proprio come il nostro rispettivamente. I madrelingua lo usano poco e mi pare che il modo generalmente corretto sia di metterlo a fine frase dopo la virgola: Tom and Sam are left and right-handed, respectively.
%Personalmente userei il piu' sicuro i.e. tanto si capisce e diciamo che diverse aree usano diversi meccanismi.
unsupervised, reinforcement, and supervised learning \citep{Doya1999WhatAretheComputationsoftheCerebellumtheBasalGangliaandtheCerebralCortex,Doya2000ComplementaryRolesofBasalGangliaandCerebelluminLearningandMotorControl}.
Recently, we have proposed a theory \citep{caligiore2019super} that expands such view by proposing that, although those learning processes might be predominant in the three macro brain systems, plasticity within them mixes the three learning mechanisms.
The hypothesis proposed here focuses on the cortex and on the empirical evidence showing that dopamine reaches cortical targets \citep{WilliamsGoldmanRakic1993CharacterizationoftheDopaminergicInnervationofthePrimateFrontalCortexUsingaDopamineSpecificAntibody, niu2020receptor}.
On this basis, the hypothesis proposes that dopamine directly conveys information on the reward to the target cortices. Therefore, learning processes happening within them integrate associative (UL) and reward-based trial-and-error mechanisms (RL).
%fede: (unsupervised learning; UL) cosi' e' come scrivere (unsupervised learning; unsupervised learning)
Here we also present a computational model that actually mixes the two learning processes.
The key idea at the core of the model is to exploit the stochastic nature of the units of Boltzmann neural networks to implement the RL key noise-based search process through the REINFORCE algorithm \citep{williams1992simple}.
The author of this algorithm envisaged the link between the stochastic units used in Boltzmann neural networks and those used in REINFORCE \citep{williams1992simple}. However, the original work proposed the algorithm only to support the acquisition of actions as usually done in RL \citep{SuttonMcAllesterSinghMansour2000Policygradientmethods,sutton1998reinforcement}.
Instead, the model proposed here uses REINFORCE to learn inner representations in the neural network.
This differs from other neural network models where reward only informs the training at the output layer of the network whereas the deeper representations are updated based on biologically non-plausible error back-propagation mechanisms \citep{ArulkumaranDeisenrothBrundageBharath2017DeepReinforcementLearningaBriefSurvey,MnihKavukcuogluSilverRusuVenessBellemareGravesRiedmillerFidjelandOstrovskiPetersenBeattieSadikAntonoglouKingKumaranWierstraLeggHassabis2015Humanlevelcontrolthroughdeepreinforcementlearning}.
To our knowledge, the proposed model is the first to allow reward signals to \textit{directly} bias the acquisition of the representations encoded in the inner neural layers of the network in a bio-plausible manner.
The model thus represents a new tool to investigate the potential utility of this for computational purposes and also to study the possible effects of the mixed UL/RL processes within the brain cortex.

%...then, passing to the results...

%Intro to model test and results.
The model was tested with a task having the features of experiments used in the field of category learning, in particular requiring sorting images based on different attributes of a given category \citep{HananiaSmith2010SelectiveAttentionandAttentionSwitchingTowardsaUnifiedDevelopmentalApproach}.
This was done to start to consider how the model could be used to study specific phenomena in category learning \citep{ZeithamovaMackBraunlichDavisSegerKesterenWutz2019BrainMechanismsofConceptLearning}, as further discussed below.

% 1) U-SHAPE: A balanced mix of UL and RL is best for performance
%
%Fig. 6,  performance of models with different RL levels
%Table 1: statistical differences.
%Main result: inverted U shape of performance
%Discussion:
The main result of the tests of the model is that a suitably balanced mix of UL and RL leads the model to achieve the best performance.
This result holds for all tested conditions, as shown in Figure~\ref{Figure:MAX_R} and Table~\ref{Table:Post_Hoc}.
Further analyses explained the possible causes of this.

% 2) PREVAILING ASSOCIATIVE LEARNING. With only associative learning the system learns faster, but then it possibly waste computational resources (which must always be limited due to the richness of sensory inputs) and remains with scarse resources for what is relevant for action
%
%Fig. 5 training curves.
%Main result: different learning speed (aside final performance) for different levels of RL
%Discussion: associative learning allows learning to take off earlier as it can work even with the erratic behaviours of RL; however, but then possibly waste computational resources and remain with scarse resources for what is relevant for action
The case of absent or low RL has some initial advantages at the beginning of training.
This can be seen from Figure~\ref{Figure:Training_Curves} where the pure UL case exhibits the learning curve with the sharpest initial increase of performance. Figures S1, S2 and S3 in Supplementary Materials show that this effect is also robustly present in sub-groups of simulations.
The reason for this is that initially the models with high RL produce a highly variable exploratory behaviour, and thus the resulting reward signals that guide the learning process involving the deeper layers of the network are rare and unreliable.
Instead, since the initial phases of training the UL process can proceed independently of the success of behaviour, and so it can build representations needed to support the learning of behaviour itself. 
However, with the advancement of training the conditions with absent/low levels of RL achieve a lower performance than the more balanced conditions, as shown in Figure~\ref{Figure:MAX_R} and Table~\ref{Table:Post_Hoc}.
Figures S1, S2 and S3 in Supplementary Materials confirm the generality of this result.
The reason is that UL tends to encode all features of the images, and so the task-relevant features compete with them and remain with insufficient computational resources. 

% 3) HIGHER RESOURCES, LESS IMPAIRING USUPERVISED LEARNING. With more resources, the model can afford more unsupervised learning
%
%Table 2. Different resource quantities.
%Main result: with more resources, the system can afford lower RL.
%Discussion: with more resources, the system can afford lower RL. In ecological conditions, informatoin is always overwhelming with respect to the available computational resources, so some task-based feature selectoin is always advantageous.
This interpretation is corroborated by the tests where we manipulated the computational resources that were available to the system, and specifically the number of the reward-biased units at the second layer of the DBN (Table~\ref{Table:Resources_Categories_conditions}).
These tests show that with a higher amount of computational resources the best performance is achieved by the models having a better balance of UL and RL, in particular a higher level of UL.
This is because the encoding of task-irrelevant features is less impairing for the encoding of task-relevant features.
To appreciate the relevance of this result, it should be considered that in ecological conditions the information received from sensors (e.g., the information from the retina) always overwhelms the available computational resources, and so some task-based feature selection is always advantageous.

% 4) HIGH RL: With low or absent unsupervised learning, RL the system learns slower; it also incurs in local minima
%
%Results: Figure 9: local minima pointed by the arrow. 
%Results: 
%Discussion: RL can lead to local minima.
At the opposite side of the spectrum, also models where RL is predominant or exclusive have computational limitations, as shown in Figure~\ref{Figure:MAX_R} and Table~\ref{Table:Post_Hoc}.
% First, plasticity based only on RL slowly than when it also involves UL because the acquisition of internal representations is guided only by reward and initially this is attained in an erratic fashion, as also discussed above (Figure~\ref{Figure:Training_Curves}).
%sotto o metti first e poi second invece di moreover e cancelli l'importantly oppure togliamo il first e la seconda frase l'apriamo con Inportantly/Crucially,...
The acquisition of internal representations is slow in models solely based on RL plasticity as learning is guided only by reward. Hence, the reconfiguration of the synaptic strengths are initially attained in an erratic fashion, as also discussed above (Figure~\ref{Figure:Training_Curves}).
Importantly, the system tends to incur in local minima with the progression of learning, as shown in Figure~\ref{Figure:Internal_Size}E.

These results suggest the general possibility that UL and RL might express advantages at different stages of learning, in particular, UL might be more useful at the beginning of learning while RL at later stages.
Future work might thus aim to study how to \textit{dynamically regulate} the UL/RL balance during learning.  

% 5) IMAGE CLUSTERING. Regarding the nature of the representations that emerge: RL leads to cluster items based on what is needed for action. This agrees with what is also shown by all standard RL models using deep nets. However, we also show how a contrary tendency can be given by unsupervised learning.
%
%Fig. 7, 8 , 9: clusters of internal representations, based on reconstruction.
%Main result: demonstration that RL leads to cluster items based on what is needed for action (as other RL models); instead, unsupervised learning leads to represent all features similarly. 
%Discussion: only models like ours, that mix the two, allow balancing the two.
%
% 6) AMOUNT OF INFORMATION DISCARDED. Unsupervised learning stores all info; RL stores less information.
%
%Fig. 10: reconstruction error caused by different UL-RL mix.
%Main result: UL discards less information, RL discards more information
%Discussion: UL stores more info; RL stores less information: likely, this allows RL to have more resources available for task solution.
%
The generative nature of the perceptual component of the model allowed the investigation of how the internal representations tend to cluster the images depending on the categorisation rule of the task and the UL/RL mix balance.
Graphs `A-B' of Figures~\ref{Figure:Internal_Colour}, \ref{Figure:Internal_Shape}, and \ref{Figure:Internal_Size} show that when RL is absent or low, UL tends to lead to the acquisition of all features of the image independently of their relevance for the task, as also shown by the low reconstruction error obtained in this cases (Figure~\ref{Figure:Min_REc_Error}).
Instead, graphs `C, D' of the same figures show how a mix of the two learning processes leads to the encoding of different attributes of the visual shapes. 
The images are grouped in relation to the responses to be associated to them, thus facilitating downstream action selection.
Finally, when learning is only driven by reward, as shown by the graphs `E' of the figures, the internal representations collapse to only one representation per group of images requiring a given action.
This interpretation is also supported by the higher levels of the reconstruction error obtained in these cases (Figure~\ref{Figure:Min_REc_Error}).

% 7) In particular, RL tends to preserve only information on key features needed for the task at hand: in the extreme case, only the key elements of images needed to solve the task.
%
% Fig. 11:  internal representations based on image reconstruction.
%Main result: >RL leads to focus on fewer aspects (even few pixels/details, in the extreme case) of images
%Discussion: higher RL leads to create more detailed representations; this favours action-oriented representations; but make system vulnerable to local minima; we expect that this leads also leads to create internal representations that generalise very poorly to new situations, as we have started to see with a preliminary investigation on cross-category experiments (data not reported). 
Figure~\ref{Figure:Reconstructions} exploits the generativity of the model to highlight the features of the images that RL tends to isolate when it is increasingly strong. 
As the figure shows, these are few features that strongly differentiate the shapes into the desired categories, for example features related to specific colours, or specific elements of the shapes or size.
These representations favour learning of the downstream actions but imply less robustness and make the system vulnerable to local minima, as we have seen above.
Moreover, this is expected to create representations that generalise poorly to new tasks, an important issue that deserves further investigation as it might justify why the brain seems to rely less on reward in early visual processing stages.

%Category learning: similar data
These results show how the presented model has the potential to be used to interpret the empirical experiments investigating the well-known phenomenon for which the tasks accomplished tend to modulate the acquired perceptual representations.
This phenomenon has been shown in mice \citep{poort2015learning}, primates \citep{sigala2002visual,de2008effects, emadi2014behavioral}, and also humans \citep{de2006discrimination, astafiev2004extrastriate}.
In particular, the studies on humans involve the wide research field of `category learning' (for reviews, see \citealt{AshbyMaddox2011HumanCategoryLearning20,ZeithamovaMackBraunlichDavisSegerKesterenWutz2019BrainMechanismsofConceptLearning}).  
The bulk of the research in this field has traditionally focused on the contrast between explicit versus procedural mechanisms for category learning \citep{maddox2004dissociating,SegerMiller2010Categorylearninginthebrain} and more recently on the prototype versus example-based nature of the acquired sensory representations \citep{MackPrestonLove2013DecodingtheBrainsAlgorithmforCategorizationfromItsNeuralImplementation,BowmanZeithamova2018AbstractMemoryRepresentationsintheVentromedialPrefrontalCortexandHippocampusSupportConceptGeneralization,ZeithamovaMackBraunlichDavisSegerKesterenWutz2019BrainMechanismsofConceptLearning}.
%
%Category representations and reward are scantly studied
Although reward and RL are considered important for category learning \citep{SegerPeterson2013CategorizationDecisionMakingGeneralization,chelazzi2013rewards}, only recently few works have started to investigate how reward affects the acquired representations, the issue relevant for the hypothesis and model proposed here.
For example, it has been shown that category learning in the ventromedial prefrontal cortex, inferior parietal cortex, and intraparietal sulcus are affected by the reward \citep{BraunlichSeger2016CategoricalEvidenceConfidenceandUrgencyduringProbabilisticCategorization,ZeithamovaMackBraunlichDavisSegerKesterenWutz2019BrainMechanismsofConceptLearning}.
%
%Current knowledge gap that we face.
To our knowledge, however, we still do not know how reward might influence the formation of low-level representations of category features, for example within the extrastriate cortex of the brain parietal areas.

%Embodiement of our model (move to the discussion)
A final remark is that the proposed model is coherent with the theoretical framework of \textit{embodied perception} \citep{DiFerdinandoParisi2004InternalRepresentationsofSensoryInputReflecttheMotorOutputwithWhichOrganismsRespondtotheInput,vernon2008cognitive,foglia2013embodied} proposing that the brain constructs internal representations of the world ``for being ready to act''. 
In this respect, the model specifies a possible ways in which the brain, based on the reward signals directly reaching its inner areas, might `warp' representations in favour of the pursued tasks.

%***********************************
\section{Conclusions}

We have proposed the hypothesis that also non-motor cortices, in particular the extra-striate cortices, learn through both associative mechanisms (unsupervised learning) and reward-based mechanisms (reinforcement learning).
Moreover, we have proposed a bio-plausible computational model facing a category-based sorting task to start to study how these mixed learning processes might affect the acquired representations of stimuli.

% The results obtained with the tests of the model can be summarised as follows:
% \begin{itemize}
%   \item 
%     A suitably balanced mix of unsupervised and RL processes leads to the highest performance.    
%   \item 
%     If the contribution of associative learning contribution is excessive, then learning is initially faster because the deep representations can form even with low reward; however, in later phases the system might have insufficient computational resources to represent the task-relevant features. 
%     In ecological conditions, the latter case is highly likely due to the fact that sensory information is always overwhelming with respect to the available computational resources.
%   \item 
%     If the contribution of RL is excessive, then learning is initially slower, can incur in local minima, and the system might also have difficulty to generalise to other tasks.
%   \item 
%     Due to the effect of reward-based RL, percepts requiring the same actions tend to develop representations that are clustered between them with respect to percepts requiring different actions.    
%   \item 
%     A higher contribution of RL tends to generate representations that preserve a smaller amount of information on percepts, and to focus on few features needed to support the selection of the task actions.
% \end{itemize}

The results obtained with the tests of the model show that a suitably balanced mix of unsupervised and reinforcement learning processes leads to the highest performance. 
On one hand, excessive unsupervised learning tends to use computational resources to represent all input features and thus to leave scarce resources for the representation of task-relevant features.
On the other hand, an excessive RL tends to lead to initial slow learning and to incur in local minima.
Moreover, the results show how reward might lead to the acquisition of action-oriented representations on the basis of bio-plausible mechanisms, and this favours the selection of downstream actions.    

In future work, the model could be used to address specific data on the reward-based modulation of category learning in the brain cortex.
Moreover, the model prompts further computational studies directed to investigate the possible advantages of reward signals directly reaching the deep layers of artificial neural networks.

\section{Acknowledgements}
We thank the Neuroscience Gateway \citep{sivagnanam2013introducing} used to run most of the simulations.
This work has received funding from the European Union’s Horizon 2020 Research and Innovation Program, under Grant Agreement No 713010 of the project `GOAL-Robots -- Goal-based Open-ended Autonomous Learning Robots', and under Grant No 796135 of the H2020-MSCA-IF-2017 project `INTENSS'. 

%Bibliography
\bibliography{References.bib}

\newpage

\onecolumn

\newcommand{\beginsupplement}{%
        \setcounter{table}{0}
        \renewcommand{\thetable}{S\arabic{table}}%
        \setcounter{figure}{0}
        \renewcommand{\thefigure}{S\arabic{figure}}%
     }

\beginsupplement

\section*{\textit{Supplementary Material}}

\subsection*{Methods: further details on the model simulations}

We tested the model solving the sorting task with different task conditions (sorting rule, i.e. colour, shape or size) and perceptual component configurations (the number of neurons of top hidden layer and the reward contribution into the learning process). We randomly changed these parameters, keeping the others fixed. Table~\ref{Table:Parameters} shows the key parameters of simulations.

\begin{table*}[htb!]
\centering
\textbf{Simulations parameters} \par\medskip
\resizebox{ \textwidth}{!}{
\begin{tabular}{|c | c | c|}
\hline
 \textbf{\thead{Label}} & \textbf{\thead{Value/Range}} & \textbf{\thead{Description}}  \\
    \hline
    Sorting rule & $\{colour, shape, size\}$ & \thead{Variable. \\ latent rule to solve the sorting task} \\
    \hline
    Training epochs & $15 * 10 \textsuperscript{3}$ & \thead{Fixed. \\ Training epochs of sorting task.} \\
     \hline
    Single-layer perceptron output units & $10$ & \thead{Fixed. \\ Output neurons of motor component.} \\
    \hline
    Single-layer perceptron learning rate (REINFORCE) & $1 * 10 \textsuperscript{-2}$ & \thead{Fixed. \\ Training learning rate of motor component.} \\
    \hline
    Multi-layers perceptron hidden units & $50$ & \thead{Fixed. \\ Hidden neurons of predictor component.} \\
    \hline
     Multi-layers perceptron learning rate (Backpropagation) & $1 * 10 \textsuperscript{-3}$ & \thead{Fixed. \\ Training learning rate of predictor component.} \\
    \hline
    Single-layer perceptron learning rate (REINFORCE) & $1 * 10 \textsuperscript{-2}$ & \thead{Fixed. \\ Training learning rate of motor component.} \\
    \hline
    DBN units (visible layer) & $2352$ & \thead{Fixed. \\ Neurons of visible layer.}\\
    \hline
    DBN units (first hidden layer) & $200$ & \thead{Fixed. \\ Neurons of first hidden layer.} \\
    \hline
    DBN units (second hidden layer) & $\{10, 50\}$ & \thead{Variable. \\ Neurons of second hidden layer.} \\
    \hline
    First RBM (off-line) training epochs & $1 * 10 \textsuperscript{3}$ & \thead{Fixed. \\ Training epochs necessary to achieve a \\ dataset reconstruction error of $0.001$.} \\
    \hline
    First RBM learning rate (Constrastive Divergence) & $1 * 10 \textsuperscript{-2}$ & \thead{Fixed. \\ Training (offline) learning rate.} \\
    \hline
    First RBM momentum (Constrastive Divergence) & $0.9$ & \thead{Fixed. \\ Training (offline) momentum.} \\
    \hline
     Second RBM learning rate (Constrastive Divergence) & $1 * 10 \textsuperscript{-3}$ & \thead{Fixed. \\ Training learning rate.} \\
    \hline
    Second RBM momentum (Constrastive Divergence) & $0.9$ & \thead{Fixed. \\ Training momentum.} \\
    \hline
     Second RBM learning rate (REINFORCE) & $1 * 10 \textsuperscript{-2}$ & \thead{Fixed. \\ Training learning rate.} \\
    \hline
     $\lambda$ & $\{1, 0.1, 0.01, 0.001, 0\}$ & \thead{Variable. \\ Contribution of the Contrastive Divergence to the weights update} \\
    \hline
     Second RBM reward contribution & \thead{$(1 - \lambda)$, \\ with $\lambda \in \{1, 0.1, 0.01, 0.001, 0\}$} & \thead{Variable. \\ Contribution of the REINFORCE to the weights update} \\
    \hline

\end{tabular}
}
\caption{The table shows the simulations parameters.}
\label{Table:Parameters}
\end{table*}

\subsection*{Results: further statistical analysis}

\subsubsection*{Performances analysis}

\begin{figure*}
    \centering
    \textbf{Learning curves of models: colour category, low computational resources} \par\medskip
	\includegraphics[width=0.8\textwidth, height = 18EM]{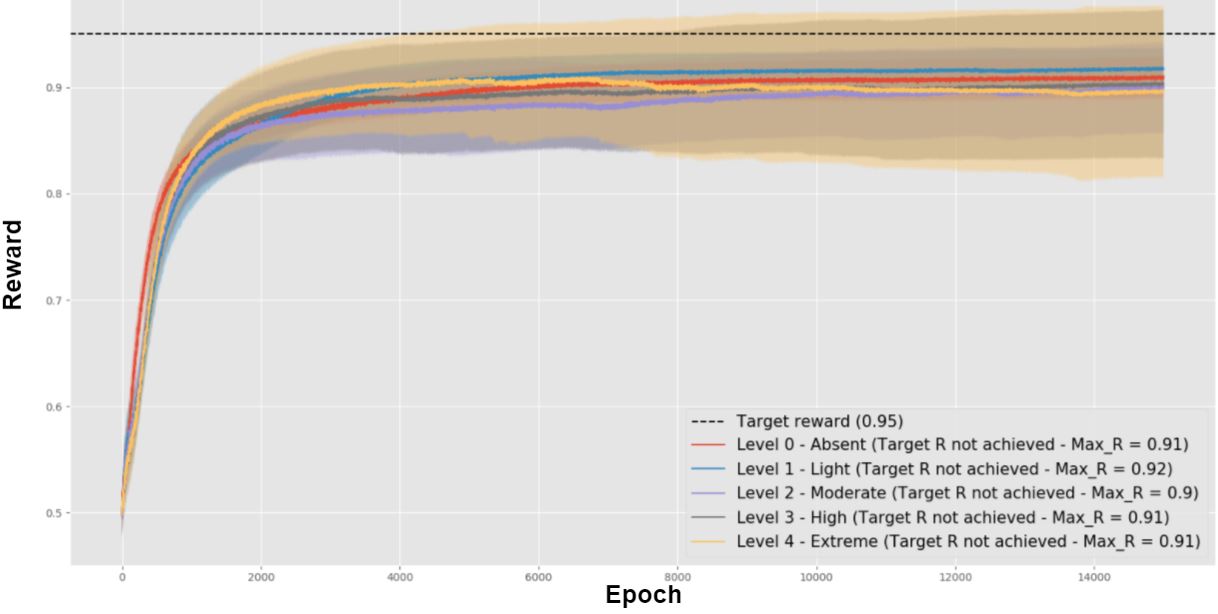}
	 \caption{Reward per epoch in the task task involving the colour category and low computational resources, of the five models involving different UL/RL levels, averaged over the models using a given level.
	 Shaded areas represent the curves standard deviations.}
	 \label{Figure:Training_Curves_Colour10}
\end{figure*}

\begin{figure*}
    \centering
    \textbf{Learning curves of models: shape category, low computational resources} \par\medskip
	\includegraphics[width=0.8\textwidth, height = 18EM]{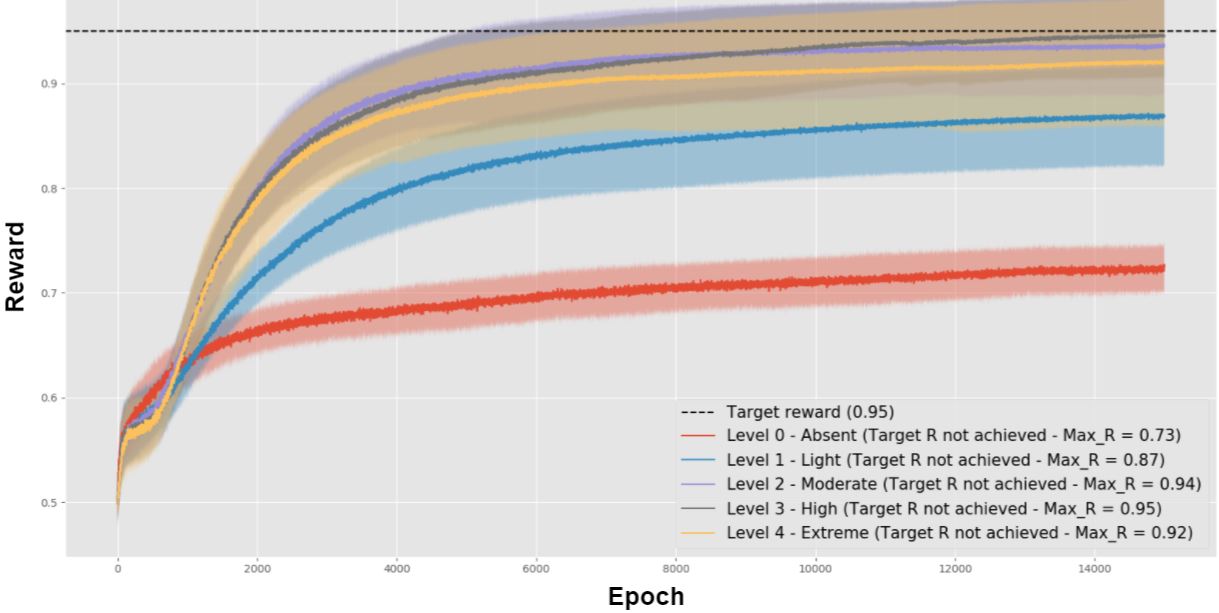}
	 \caption{Reward per epoch in the task task involving the shape category and low computational resources, of the five models involving different UL/RL levels, averaged over the models using a given level.
	 Shaded areas represent the curves standard deviations.}
	 \label{Figure:Training_Curves_Shape10}
\end{figure*}

\begin{figure*}
    \centering
    \textbf{Learning curves of models: size category, low computational resources} \par\medskip
	\includegraphics[width=0.8\textwidth, height = 18EM]{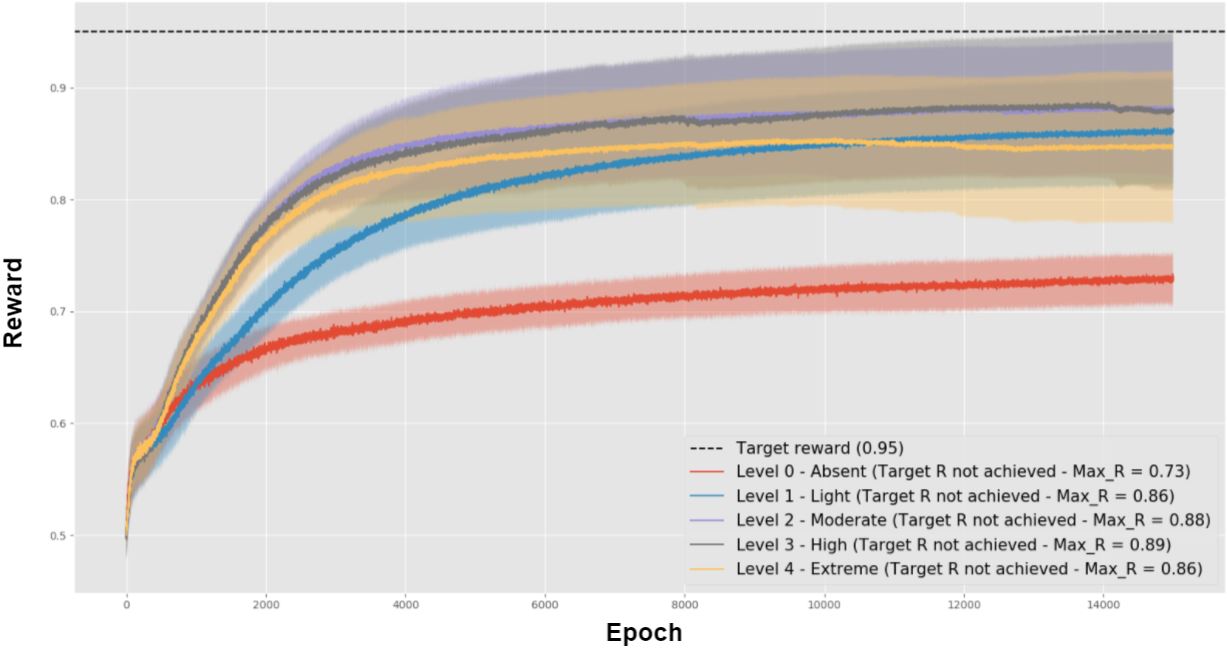}
	 \caption{Reward per epoch in the task task involving the size category and low computational resources, of the five models involving different UL/RL levels, averaged over the models using a given level.
	 Shaded areas represent the curves standard deviations.}
	 \label{Figure:Training_Curves_Size10}
\end{figure*}

\begin{figure*}
    \centering
    \textbf{Learning curves of models: colour category, high computational resources} \par\medskip
	\includegraphics[width=0.8\textwidth, height = 18EM]{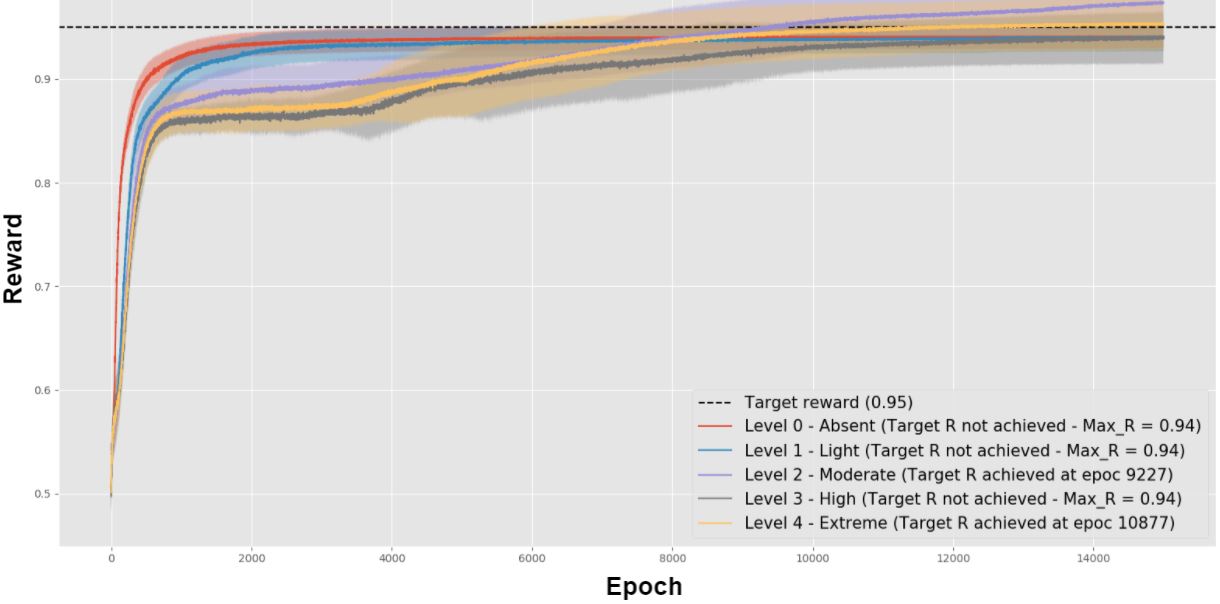}
	 \caption{Reward per epoch in the task task involving the colour category and high computational resources, of the five models involving different UL/RL levels, averaged over the models using a given level.
	 Shaded areas represent the curves standard deviations.}
	 \label{Figure:Training_Curves_Colour}
\end{figure*}

\begin{figure*}
    \centering
    \textbf{Learning curves of models: shape category, high computational resources} \par\medskip
	\includegraphics[width=0.8\textwidth, height = 18EM]{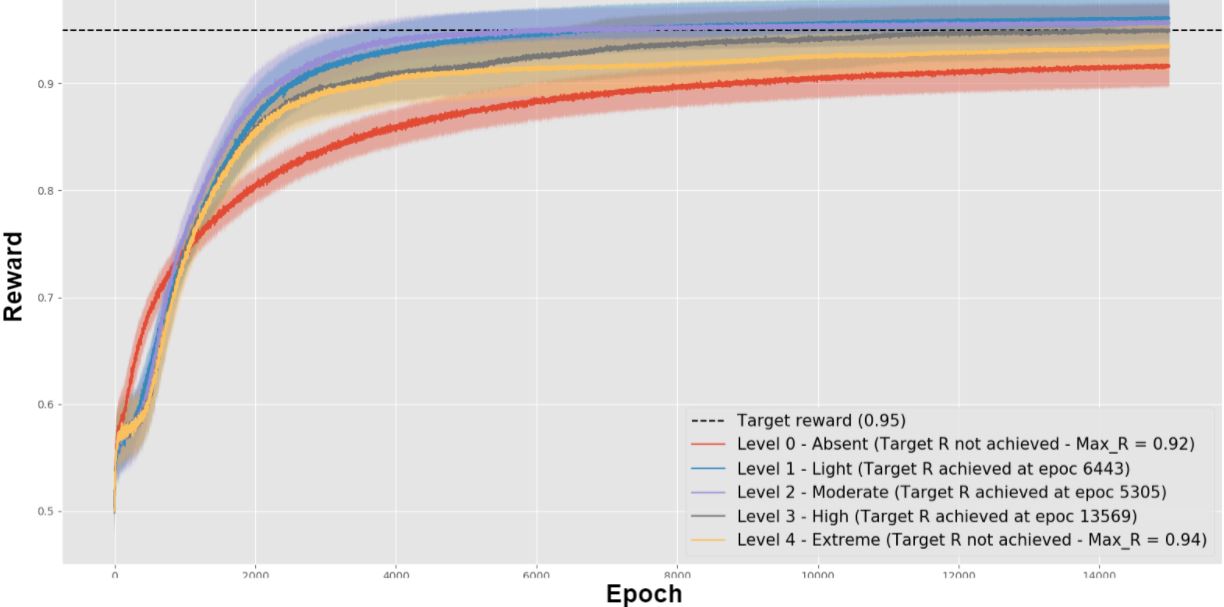}
	 \caption{Reward per epoch in the task task involving the shape category and high computational resources, of the five models involving different UL/RL levels, averaged over the models using a given level.
	 Shaded areas represent the curves standard deviations.}
	 \label{Figure:Training_Curves_Shape}
\end{figure*}

\begin{figure*}
    \centering
    \textbf{Learning curves of models: size category, high computational resources} \par\medskip
	\includegraphics[width=0.8\textwidth, height = 18EM]{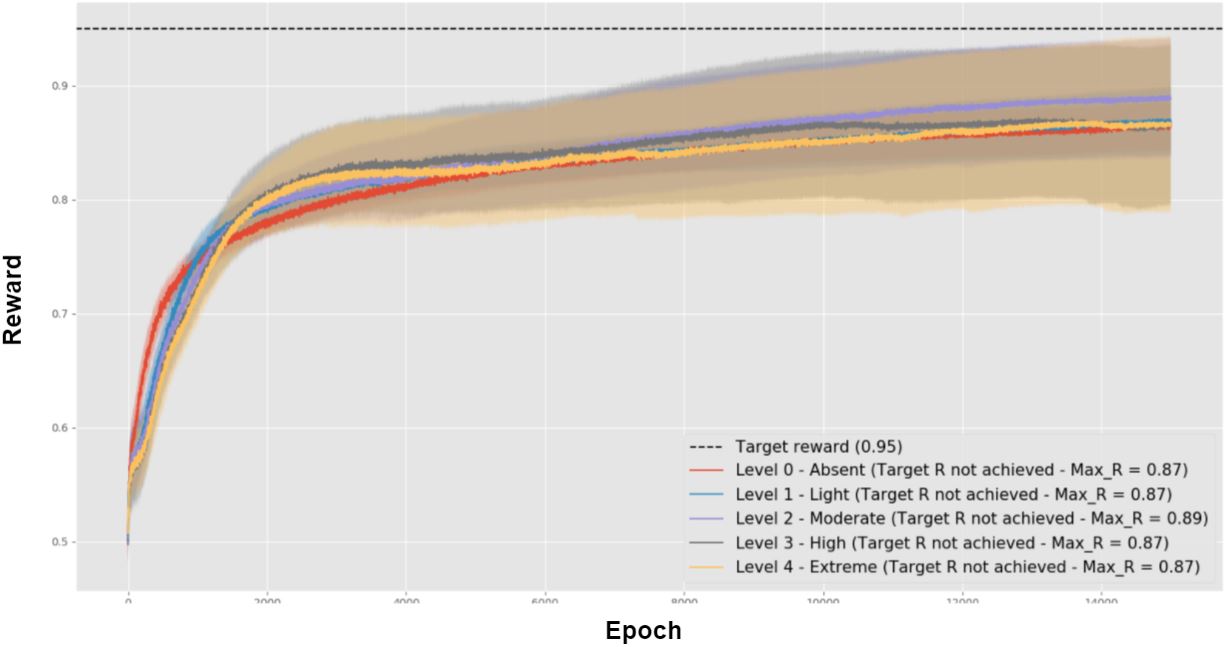}
	 \caption{Reward per epoch in the task task involving the size category and high computational resources, of the five models involving different UL/RL levels, averaged over the models using a given level.
	 Shaded areas represent the curves standard deviations.}
	 \label{Figure:Training_Curves_Size}
\end{figure*}

%Training curves
Figures 
\ref{Figure:Training_Curves_Colour10},
\ref{Figure:Training_Curves_Shape10}, 
 and
\ref{Figure:Training_Curves_Size10} show the training curves of the models trained with different RL contributions in 15,000 epochs, for the three sorting categories and for the condition with 10 units of the DBN second hidden layer.
Figures 
\ref{Figure:Training_Curves_Colour},
\ref{Figure:Training_Curves_Shape}, and
\ref{Figure:Training_Curves_Size} show the analogous curves for the condition involving 50 units of the DBN second hidden layer.
In particular, these are the models of the condition with a high-level of computational resources, namely 50 units at the level of the second hidden layer of the DBN.
In all conditions, the L0 models (with no reinforcement learning - RL, i.e. relying only on  unsupervised learning - UL) show an initial highest performance with respect to the other models L1, L2, L3, and L4.
This confirms that in L0 models representation learning is initially facilitated with respect to models with a higher RL contribution as the reward is initially erratic.
Moreover, for all three category tasks  the reward achieves a maximum final performance for the L2 models having a balanced level of UL and RL.
Indeed, these models outperform the models with absent or very low RL (L0 and L1) because these employ a lot of computational resources for non-task specific features; moreover they outperform the models with very high or extreme RL (L3 and L4) because these tend to incur in local minima.

%10 hidden units


%50 hidden units


Table~\ref{Table:Colour_low} shows the post-hoc tests with the Bonferroni correction.
The tests are grouped for each specific combination of the three main conditions, that is, computational resources (2 conditions), sorting rule used in the task (3 conditions), and reward contribution (5 conditions).
\begin{table*}[htb!]
\centering
\textbf{Sorting rule: Colour, Computational Resources: Low} \par\medskip
\vspace{5mm}
\resizebox{ \textwidth}{!}{
\begin{tabular}{|c | c | c | c | c | c|}
\hline
 & \textbf{Absent (L0, N = $40$)} & \textbf{Low (L1, N = $34$)} & \textbf{Moderate (L2, N = $21$)} & \textbf{High (L3, N = $38$)} & \textbf{Extreme (L4, N = $32$)} \\
    \hline
   \textbf{Absent (L0)}        & //  & $p > 0.05$ (NS) & $p > 0.05$ (NS) & $p > 0.05$ (NS) & $p > 0.05$ (NS) \\
    \hline
    \textbf{Low (L1)}        & // & //  & $p > 0.05$ (NS) & $p > 0.05$ (NS) & $p > 0.05$ (NS) \\
    \hline
    \textbf{Moderate (L2)}       & //  & //  & // & $p > 0.05$ (NS) & $p > 0.05$ (NS) \\
    \hline
    \textbf{High (L3)}       & //  & //  & // & //  & $p > 0.05$ (NS) \\
    \hline
    \textbf{Extreme (L4)}        & //  & //  & // & // & // \\
    \hline
\end{tabular}
}
\caption{The table shows the post hoc multiple comparisons (t-test with Bonferroni correction) of models in case of the colour sorting rule and low computational resources. NS = not significant.}
\label{Table:Colour_low}
\end{table*}

\begin{table*}[htb!]
\centering
\textbf{Sorting rule: Shape, Computational Resources: Low} \par\medskip
\vspace{5mm}
\resizebox{ \textwidth}{!}{
\begin{tabular}{|c | c | c | c | c | c|}
\hline
 & \textbf{Absent (L0, N = $32$)} & \textbf{Low (L1, N = $44$)} & \textbf{Moderate (L2, N = $29$)} & \textbf{High (L3, N = $42$)} & \textbf{Extreme (L4, N = $39$)} \\
    \hline
   \textbf{Absent (L0)}        & //  & $p < 0.001$ & $p < 0.001$ & $p < 0.001$ & $p < 0.001$ \\
    \hline
    \textbf{Low (L1)}        & // & //  & $p < 0.001$ & $p < 0.001$ & $p < 0.01$ \\
    \hline
    \textbf{Moderate (L2)}       & //  & //  & // & $p > 0.05$ (NS) & $p > 0.05$ (NS) \\
    \hline
    \textbf{High (L3)}       & //  & //  & // & //  & $p > 0.05$ (NS) \\
    \hline
    \textbf{Extreme (L4)}        & //  & //  & // & // & // \\
    \hline
\end{tabular}
}
\caption{The table shows the post hoc multiple comparisons (t-test with Bonferroni correction) of models in case of the shape sorting rule and low computational resources. NS = not significant.}
\label{Table:Shape_low}
\end{table*}

\begin{table*}[htb!]
\centering
\textbf{Sorting rule: Size, Computational Resources: Low} \par\medskip
\vspace{5mm}
\resizebox{ \textwidth}{!}{
\begin{tabular}{|c | c | c | c | c | c|}
\hline
 & \textbf{Absent (L0, N = $38$)} & \textbf{Low (L1, N = $35$)} & \textbf{Moderate (L2, N = $39$)} & \textbf{High (L3, N = $28$)} & \textbf{Extreme (L4, N = $41$)} \\
    \hline
   \textbf{Absent (L0)}        & //  & $p < 0.001$ & $p < 0.001$ & $p < 0.001$ & $p < 0.001$ \\
    \hline
    \textbf{Low (L1)}        & // & //  & $p > 0.05$ (NS) & $p > 0.05$ (NS) & $p > 0.05$ (NS) \\
    \hline
    \textbf{Moderate (L2)}       & //  & //  & // & $p > 0.05$ (NS) & $p > 0.05$ (NS) \\
    \hline
    \textbf{High (L3)}       & //  & //  & // & //  & $p > 0.05$ (NS) \\
    \hline
    \textbf{Extreme (L4)}        & //  & //  & // & // & // \\
    \hline
\end{tabular}
}
\caption{The table shows the post hoc multiple comparisons (t-test with Bonferroni correction) of models in case of the size sorting rule and low computational resources. NS = not significant.}
\label{Table:Size_low}
\end{table*}

\begin{table*}[htb!]
\centering
\textbf{Sorting rule: Colour, Computational Resources: High} \par\medskip
\vspace{5mm}
\resizebox{ \textwidth}{!}{
\begin{tabular}{|c | c | c | c | c | c|}
\hline
 & \textbf{Absent (L0, N = $39$)} & \textbf{Low (L1, N = $33$)} & \textbf{Moderate (L2, N = $31$)} & \textbf{High (L3, N = $20$)} & \textbf{Extreme (L4, N = $43$)} \\
    \hline
   \textbf{Absent (L0)}        & //  & $p > 0.05$ (NS) & $p < 0.001$ & $p > 0.05$ (NS) & $p < 0.01$ \\
    \hline
    \textbf{Low (L1)}        & // & //  & $p < 0.001$ & $p > 0.05$ (NS) & $p < 0.05$ \\
    \hline
    \textbf{Moderate (L2)}       & //  & //  & // & $p < 0.001$ & $p < 0.01$ \\
    \hline
    \textbf{High (L3)}       & //  & //  & // & //  & $p > 0.05$ (NS) \\
    \hline
    \textbf{Extreme (L4)}        & //  & //  & // & // & // \\
    \hline
\end{tabular}
}
\caption{The table shows the post hoc multiple comparisons (t-test with Bonferroni correction) of models in case of the colour sorting rule and high computational resources. NS = not significant.}
\label{Table:Colour_high}
\end{table*}

\begin{table*}[htb!]
\centering
\textbf{Sorting rule: Shape, Computational Resources: High} \par\medskip
\vspace{5mm}
\resizebox{ \textwidth}{!}{
\begin{tabular}{|c | c | c | c | c | c|}
\hline
 & \textbf{Absent (L0, N = $41$)} & \textbf{Low (L1, N = $35$)} & \textbf{Moderate (L2, N = $33$)} & \textbf{High (L3, N = $29$)} & \textbf{Extreme (L4, N = $33$)} \\
    \hline
   \textbf{Absent (L0)}        & //  & $p < 0.001$ & $p < 0.001$ & $p < 0.001$ & $p > 0.05$ (NS) \\
    \hline
    \textbf{Low (L1)}        & // & //  & $p > 0.05$ (NS) & $p > 0.05$ (NS) & $p < 0.001$ \\
    \hline
    \textbf{Moderate (L2)}       & //  & //  & // & $p > 0.05$ (NS) & $p < 0.001$ \\
    \hline
    \textbf{High (L3)}       & //  & //  & // & //  & $p < 0.05$ \\
    \hline
    \textbf{Extreme (L4)}        & //  & //  & // & // & // \\
    \hline
\end{tabular}
}
\caption{The table shows the post hoc multiple comparisons (t-test with Bonferroni correction) of models in case of the shape sorting rule and high computational resources. NS = not significant.}
\label{Table:Shape_high}
\end{table*}

\begin{table*}[htb!]
\centering
\textbf{Sorting rule: Size, Computational Resources: High} \par\medskip
\vspace{5mm}
\resizebox{ \textwidth}{!}{
\begin{tabular}{|c | c | c | c | c | c|}
\hline
 & \textbf{Absent (L0, N = $24$)} & \textbf{Low (L1, N = $30$)} & \textbf{Moderate (L2, N = $35$)} & \textbf{High (L3, N = $29$)} & \textbf{Extreme (L4, N = $29$)} \\
    \hline
   \textbf{Absent (L0)}        & //  & $p > 0.05$ (NS) & $p > 0.05$ (NS) & $p > 0.05$ (NS) & $p > 0.05$ (NS) \\
    \hline
    \textbf{Low (L1)}        & // & //  & $p > 0.05$ (NS) & $p > 0.05$ (NS) & $p > 0.05$ (NS) \\
    \hline
    \textbf{Moderate (L2)}       & //  & //  & // & $p > 0.05$ (NS) & $p > 0.05$ (NS) \\
    \hline
    \textbf{High (L3)}       & //  & //  & // & //  & $p > 0.05$ (NS) \\
    \hline
    \textbf{Extreme (L4)}        & //  & //  & // & // & // \\
    \hline
\end{tabular}
}
\caption{The table shows the post hoc multiple comparisons (t-test with Bonferroni correction) of models in case of the size sorting rule and high computational resources. NS = not significant.}
\label{Table:Size_high}
\end{table*}

\subsection*{Reconstruction error and information stored}

%Gianluca: Giovanni, e' questo indicato nella frase sotto il motivo di questa sezione?
In this section we explain why the reconstruction errors of the DBN reported in the main text can be considered a measure of the information on the input patterns retained by this component of the models.
Restricted Boltzmann Machines and Deep Belief Networks are generative models able to store the joint probability between an input and the consequent hidden layer activation \citep{hinton2006fast, Hinton2012}. 
This property makes these models able to execute a dimensional reduction of input patterns \citep{hinton2006reducing} and to `generate' such input patters based on an inverse spread of activation spread from a hidden layer towards the visible layer. 
Due to the difficulty of meaningfully activating the distributed representations within the hidden layers in a direct way, a typical way to exploit this generativity property also followed here is to precede the hidden-visible activation spreading by a standard visible-hidden activation. 
This allows the computation of the \textit{reconstruction error}, corresponding to the difference between an input pattern and the corresponding reconstruction.
This error is relevant as it represents a measure of the information that the system has retained on the input pattern.

\subsection*{Internal representations analysis: PCA and K-means details}

In the main test we illustrated the results obtained on average over whole classes of simulations. 
Here we show the outcome of the PCA (principal component analysis) and K-means analyses exemplifying the results within each class.
In particular, we considered examples that were more aligned with the average scores of the classes as they should be more representative of the classes themselves.

Tables~\ref{Table:Colour_PCA},~\ref{Table:Shape_PCA}, and ~\ref{Table:Size_PCA} show the cumulative explained variance of the the PCA in correspondence to a growing number of principal components. 
The plots presented in the main text had an $n = 2$ corresponding to the first two principal components.
This value is acceptable because it is almost always higher than the median cumulative explained variance and at the same time allowed us to plot the components of the reconstructed images.
An interesting feature that emerges from the values is that with a higher value of RL the `elbow' of the curves represented by the numbers reported in the tables become sharper.
This is in line with the fact that with a higher RL contribution the images tend to be increasingly clustered into groups corresponding to the actions to be returned while the task-irrelevant features are discarded, thus needing less components to be represented.

\begin{table*}[htb!]
\centering
\large{\textbf{\thead{PCA cumulative variance explained \\ (Sorting rule: Colour)}}} \par\medskip
\vspace{5mm}
% \resizebox{\textwidth}{!}{
\begin{tabular}{|c | c | c | c | c | c|}
\hline
 & \textbf{Absent (L0)} & \textbf{Low (L1)} & \textbf{Moderate (L2)} & \textbf{High (L3)} & \textbf{Extreme (L4)} \\
    \hline
   \textbf{N = 1}        &  $0.39$ & $0.48$ & $0.67$ & $0.63$ & $0.64$ \\
    \hline
    \textbf{N = 2}        & \textbf{0.62} & \textbf{0.76}  & \textbf{0.99} & \textbf{0.99} & \textbf{0.99} \\
    \hline
    \textbf{N = 3}       & $0.74$  & $0.86$  & $1$ & $1$ & $1$ \\
    \hline
    \textbf{N = 4}       & $0.81$  & $0.91$  & $1$ & $1$ & $1$ \\
    \hline
    \textbf{N = 5}        & $0.85$  & $0.94$  & $1$ & $1$ & $1$ \\
    \hline
    \textbf{N = 6}        & $0.89$  & $0.96$  & $1$ & $1$ & $1$ \\
    \hline
    \textbf{N = 7}        & $0.91$  & $0.97$  & $1$ & $1$ & $1$ \\
    \hline
    \textbf{Median} & $0.81$ & $0.85$  & $1$ & $1$ & $1$ \\
    \hline
\end{tabular}
% }
\caption{Cumulative explained variance (CEV) of the PCA run over the reconstructed images of the models, in the case of the colour sorting rule and low computational resources.
The $n = 2$ CEV values are highlighted in bold.}
\label{Table:Colour_PCA}
\end{table*}

\begin{table*}[htb!]
\centering
\large{\textbf{\thead{PCA cumulative variance explained \\ (Sorting rule: Shape)}}} \par\medskip
\vspace{5mm}
% \resizebox{\textwidth}{!}{
\begin{tabular}{|c | c | c | c | c | c|}
\hline
 & \textbf{Absent (L0)} & \textbf{Low (L1)} & \textbf{Moderate (L2)} & \textbf{High (L3)} & \textbf{Extreme (L4)} \\
    \hline
   \textbf{N = 1}        &  $0.39$ & $0.48$ & $0.53$ & $0.73$ & $0.67$ \\
    \hline
    \textbf{N = 2}        & \textbf{0.64} & \textbf{0.74}  & \textbf{0.87} & \textbf{0.94} & \textbf{0.98} \\
    \hline
    \textbf{N = 3}       & $0.76$  & $0.83$  & $0.99$ & $0.99$ & $1$ \\
    \hline
    \textbf{N = 4}       & $0.82$  & $0.89$  & $0.99$ & $1$ & $1$ \\
    \hline
    \textbf{N = 5}        & $0.85$  & $0.93$  & $1$ & $1$ & $1$ \\
    \hline
    \textbf{N = 6}        & $0.88$  & $0.94$  & $1$ & $1$ & $1$ \\
    \hline
    \textbf{N = 7}        & $0.91$  & $0.96$  & $1$ & $1$ & $1$ \\
    \hline
    \textbf{Median} & $0.82$ & $0.89$  & $1$ & $1$ & $1$ \\
    \hline
\end{tabular}
% }
\caption{Cumulative explained variance (CEV) of the PCA run over the reconstructed images of the models, in the case of the shape sorting rule and low computational resources.
The $n = 2$ CEV values are highlighted in bold.}
\label{Table:Shape_PCA}
\end{table*}

\begin{table*}[htb!]
\centering
\large{\textbf{\thead{PCA cumulative variance explained \\ (Sorting rule: Size)}}} \par\medskip
\vspace{5mm}
% \resizebox{\textwidth}{!}{
\begin{tabular}{|c | c | c | c | c | c|}
\hline
 & \textbf{Absent (L0)} & \textbf{Low (L1)} & \textbf{Moderate (L2)} & \textbf{High (L3)} & \textbf{Extreme (L4)} \\
    \hline
   \textbf{N = 1}        &  $0.39$ & $0.47$ & $0.61$ & $0.70$ & $0.60$ \\
    \hline
    \textbf{N = 2}        & \textbf{0.63} & \textbf{0.76}  & \textbf{0.91} & \textbf{0.90} & \textbf{1} \\
    \hline
    \textbf{N = 3}       & $0.75$  & $0.86$  & $0.99$ & $0.99$ & $1$ \\
    \hline
    \textbf{N = 4}       & $0.81$  & $0.92$  & $1$ & $1$ & $1$ \\
    \hline
    \textbf{N = 5}        & $0.85$  & $0.95$  & $1$ & $1$ & $1$ \\
    \hline
    \textbf{N = 6}        & $0.88$  & $0.96$  & $1$ & $1$ & $1$ \\
    \hline
    \textbf{N = 7}        & $0.90$  & $0.97$  & $1$ & $1$ & $1$ \\
    \hline
    \textbf{Median} & $0.81$ & $0.92$  & $1$ & $1$ & $1$ \\
    \hline
\end{tabular}
% }
\caption{Cumulative explained variance (CEV) of the PCA run over the reconstructed images of the models, in the case of the size sorting rule and low computational resources.
The $n = 2$ CEV values are highlighted in bold.}
\label{Table:Size_PCA}
\end{table*}

%SILHOUETTE VALUES
Tables~\ref{Table:Colour_Sil},~\ref{Table:Shape_Sil},~\ref{Table:Size_Sil}
show the silhouette values of the k-means algorithm corresponding to different K values establishing the number of the searched classes.
The tables show that the the highest silhouette values tend to correspond to $K=4$, the value used in the analyses reported in the main text.
This value is relevant as it corresponds to the number of attributes in each category and to which the model has to assign a different action
(colour: red, green, blue, yellow; 
form: square, circle, triangle, bar; 
size: large, medium-large, medium-small, small).
It is also interesting to observe that the best silhouette value is more highly differentiated from other values in correspondence to higher levels of RL contribution: this agrees with the fact that in these conditions the model tends to encode features that are more closely related to the actions.

\begin{table*}[htb!]
\centering
\large{\textbf{\thead{K-means Silhouette values \\ (Sorting rule: Colour)}}} \par\medskip
\vspace{5mm}
% \resizebox{\textwidth}{!}{
\begin{tabular}{|c | c | c | c | c | c|}
\hline
 & \textbf{Absent (L0)} & \textbf{Low (L1)} & \textbf{Moderate (L2)} & \textbf{High (L3)} & \textbf{Extreme (L4)} \\
    \hline
   \textbf{K = 2}        &  $0.47$ & $0.50$ & $0.71$ & $0.55$ & $0.63$ \\
    \hline
    \textbf{K = 3}        & $0.56$ & $0.61$  & $0.91$ & $0.80$ & $0.80$ \\
    \hline
    \textbf{K = 4}       & \textbf{0.64}  & \textbf{0.65}  & \textbf{1} & \textbf{1} & \textbf{1} \\
    \hline
    \textbf{K = 5}       & $0.66$  & $0.63$  & $0.86$ & $0.69$ & $0.91$ \\
    \hline
    \textbf{K = 6}        & $0.69$  & $0.64$  & $0.53$ & $0.53$ & $0.78$ \\
    \hline
    \textbf{K = 7}        & $0.73$  & $0.67$  & $0.52$ & $0.27$ & $0.66$ \\
    \hline
    \textbf{K = 8}        & $0.72$  & $0.66$  & $0.44$ & $0.42$ & $0.60$ \\
    \hline
    \textbf{Mean} & $0.64$ & $0.62$  & $0.71$ & $0.61$ & $0.77$ \\
    \hline
\end{tabular}
% }
\caption{The table shows the K-means silhouette values of models in case of colour sorting rule and low computational resources. The K = 4 silhouette values are highlighted in bold}
\label{Table:Colour_Sil}
\end{table*}

\begin{table*}[htb!]
\centering
\large{\textbf{\thead{K-means Silhouette values \\ (Sorting rule: Shape)}}} \par\medskip
\vspace{5mm}
% \resizebox{ \textwidth}{!}{
\begin{tabular}{|c | c | c | c | c | c|}
\hline
 & \textbf{Absent (L0)} & \textbf{Low (L1)} & \textbf{Moderate (L2)} & \textbf{High (L3)} & \textbf{Extreme (L4)} \\
    \hline
   \textbf{K = 2}        &  $0.44$ & $0.51$ & $0.55$ & $0.69$ & $0.63$ \\
    \hline
    \textbf{K = 3}        & $0.53$ & $0.61$  & $0.74$ & $0.84$ & $0.82$ \\
    \hline
    \textbf{K = 4}       & \textbf{0.63}  & \textbf{0.68}  & \textbf{0.94} & \textbf{0.99} & \textbf{0.92} \\
    \hline
    \textbf{K = 5}       & $0.65$  & $0.66$  & $0.99$ & $0.98$ & $0.98$ \\
    \hline
    \textbf{K = 6}        & $0.67$  & $0.66$  & $0.98$ & $0.96$ & $0.98$ \\
    \hline
    \textbf{K = 7}        & $0.70$  & $0.66$  & $0.80$ & $0.93$ & $0.71$ \\
    \hline
    \textbf{K = 8}        & $0.66$  & $0.63$  & $0.80$ & $0.94$ & $0.65$ \\
    \hline
    \textbf{Mean} & $0.61$ & $0.63$  & $0.83$ & $0.90$ & $0.81$ \\
    \hline
\end{tabular}
% }
\caption{The table shows the K-means silhouette values of models in case of shape sorting rule and low computational resources. The K = 4 silhouette values are highlighted in bold}
\label{Table:Shape_Sil}
\end{table*}

\begin{table*}[htb!]
\centering
\large{\textbf{\thead{K-means Silhouette values \\ (Sorting rule: Size)}}} \par\medskip
\vspace{5mm}
% \resizebox{ \textwidth}{!}{
\begin{tabular}{|c | c | c | c | c | c|}
\hline
 & \textbf{Absent (L0)} & \textbf{Low (L1)} & \textbf{Moderate (L2)} & \textbf{High (L3)} & \textbf{Extreme (L4)} \\
    \hline
   \textbf{K = 2}        &  $0.47$ & $0.53$ & $0.65$ & $0.70$ & $0.75$ \\
    \hline
    \textbf{K = 3}        & $0.55$ & $0.63$  & $0.87$ & $0.83$ & $0.86$ \\
    \hline
    \textbf{K = 4}       & \textbf{0.64}  & \textbf{0.72}  & \textbf{0.95} & \textbf{1.0} & \textbf{0.99} \\
    \hline
    \textbf{K = 5}       & $0.64$  & $0.71$  & $0.97$ & $0.87$ & $0.98$ \\
    \hline
    \textbf{K = 6}        & $0.66$  & $0.71$  & $0.94$ & $0.71$ & $0.69$ \\
    \hline
    \textbf{K = 7}        & $0.68$  & $0.72$  & $0.97$ & $0.48$ & $0.66$ \\
    \hline
    \textbf{K = 8}        & $0.68$  & $0.72$  & $0.95$ & $0.48$ & $0.49$ \\
    \hline
    \textbf{Mean} & $0.62$ & $0.68$  & $0.90$ & $0.73$ & $0.78$ \\
    \hline
\end{tabular}
% }
\caption{The table shows the K-means silhouette values of models in case of size sorting rule and low computational resources. The K = 4 silhouette values are highlighted in bold}
\label{Table:Size_Sil}
\end{table*}

\end{document}